\documentclass[11pt]{article}
\usepackage{authblk}
\usepackage{setspace}
\pdfoutput=1

\usepackage[margin=2.5cm]{geometry}

\usepackage{amsmath}
\usepackage{amsfonts}
\usepackage{amssymb}
\usepackage{amsthm}
\usepackage{mathtools}
\usepackage{physics}
\usepackage{siunitx}

\usepackage{adjustbox}
\usepackage{epstopdf}
\usepackage{float}
\usepackage{graphicx}
\usepackage{rotating}
\usepackage{subcaption} 
\usepackage{tikz}

\usepackage{booktabs}
\usepackage{longtable}
\usepackage{threeparttable}
\usepackage{multirow}

\usepackage{xcolor}
\usepackage[ruled]{algorithm2e}
\usepackage[labelfont={bf}]{caption}
\usepackage{enumerate}
\usepackage{lastpage}
\usepackage{listings}
\usepackage{ulem}
\usepackage{hyperref}
\usepackage{eurosym}

\usepackage{natbib}
\usepackage{verbatim}
\usepackage{url}
\usepackage{etoolbox}

\usepackage{draftwatermark}
\SetWatermarkText{}  

\newcommand{\red}[1]{\textcolor{red}{#1}}

\providetoggle{showpt}
\settoggle{showpt}{false}

\newcommand{\R}{\mathbb{R}}

\date{\today} 
\allowdisplaybreaks[1]

\newenvironment{conditions*}
{\par\vspace{\abovedisplayskip}\noindent
	\tabularx{\columnwidth}{>{$}l<{$} @{${}={}$} >{\raggedright\arraybackslash}X}}
{\endtabularx\par\vspace{\belowdisplayskip}}

\usepackage{setspace}
\begin{document}

\normalem 




\title{STraM: A strategic network design model for national freight transport decarbonization}


\author[1]{Steffen J.S. Bakker\thanks{Corresponding author: \texttt{steffen.bakker@ntnu.no}}}
\author[1,3]{Jonas Martin}
\author[2,1]{E. Ruben van Beesten}
\author[1]{Ingvild Synnøve Brynildsen}
\author[1]{Anette Sandvig}
\author[1]{Marit Siqveland}
\author[3]{Antonia Golab}

\affil[1]{Department of Industrial Economics and Technology Management, Norwegian University of Science and Technology, Alfred Getz veg 3, NO-7491 Trondheim, Norway}
\affil[2]{Econometric Institute, Erasmus University Rotterdam, Burgemeester Oudlaan 50, Rotterdam 3062 PA, the Netherlands}
\affil[3]{Energy Economy Group (EEG), Technische Universität Wien, Gusshausstraße 25-29, E370-3, Vienna,1040, Austria}
\affil[4]{Institute of Future Fuels, German Aerospace Center (DLR), 51147 Cologne, Germany}

\maketitle

\begin{abstract}
National freight transport models are valuable tools for assessing the impact of various policies and investments on achieving decarbonization targets under different future scenarios. However, these models struggle to address several critical elements necessary for strategic planning, such as the development and adoption of new fuel technologies over time, inertia in transport fleets, and uncertainty surrounding future transport costs. 
In this paper, we develop a strategic network design model, named STraM, that explicitly incorporates these key factors. STraM provides a network design plan that includes infrastructure investments and fuel technology decisions, aiming to achieve cost-effective decarbonization of the freight transport system. The model's output can be used as input for higher-resolution national freight transport models to yield results with greater operational detail. We demonstrate the application of STraM through a case study of Norway, offering valuable insights into the strategic planning of sustainable freight transport.

\end{abstract}



\section{Introduction} \label{sec:introduction}

In the context of the energy transition, 
decarbonizing freight transport is an important step to achieve emission targets set by national governments \citep{Meyer2020DecarbonizingAnalysis}. This process requires a shift from fossil to renewable fuels and possibly from higher to lower emission transport modes.  
To facilitate these shifts, governments can implement policies that focus on specific infrastructure investments, such as the construction of new roads or railways and new charging or filling stations, the expansions of terminal capacities, or enhancements to existing infrastructure. Additionally, fiscal measures, like fuel pricing or road tolls, and regulatory measures, like limits on truck size and weight, can further promote these changes \citep{Nassar2023ATransportation}. 

However, making decisions about what policies to implement -- especially those involving significant infrastructure investments -- is challenging due to the high uncertainty regarding future developments.
This raises a critical question for national governments:
which policies and infrastructure investments will most effectively contribute to achieving their desired decarbonization targets \citep{Kaack2018DecarbonizingShift}?
To answer this question, various decision support tools can be employed.


The decision support tools most commonly used by governments are so-called national freight transport models \citep{deJong2013RecentEurope}. These models describe the national freight transport system with a high level of granularity, including a detailed description of transport demand of various product types, transport infrastructure, different types of transport modes and vehicles, and many other logistics elements. These models are very useful for analyzing the impacts of proposed policy measures or specific infrastructure investments in a given future scenario. 

However, a drawback of national freight transport models is that they are \textit{static} (one-period) and \textit{deterministic}. This is problematic in a setting where \textit{strategic} decisions need to be made that have long-term effects that significantly depend on the uncertain outcomes of future events. The problem of decarbonizing the freight transport system is precisely such a setting: policy measures and infrastructure decisions have to be taken today, while their effects take place over the coming decades and depend heavily on uncertain future factors, such as the development of new fuel technologies.

In this paper, we develop a decision support tool for decarbonizing national freight transport, called STraM (\textbf{S}trategic \textbf{Tra}nsport \textbf{M}odel), that is tailored to answer \textit{strategic} questions in an uncertain system. 
STraM is complementary to the existing high-resolution national freight transport models. We sacrifice some operational detail, but in return we explicitly incorporate important elements for strategic decision-making, such as uncertainty over a long time horizon, while retaining a spatial representation of the freight transport network. The output of STraM includes a set of infrastructure investments designed to minimize the total (risk-corrected) expected system cost. Additionally, it provides an assignment of transport demand to routes across the transportation network, specifying the modes and fuels to be used.

STraM is based on the multimodal transport network design modeling framework STAN \citep{Crainic1990StrategicSystem}. 
To this starting point, we add two aspects that have not traditionally been included in national freight transport models, but that are crucial for strategic planning of transport systems decarbonization.
First, STraM encompasses \textit{multiple strategic time periods}. This allows for an accurate representation of the development of elements such as: the development and adoption of new technologies, inertia in transport fleets, a decreasing emission budget, and investments over time. Second, STraM explicitly models \textit{long-term uncertainty} in cost and technology development. 
This is especially relevant in the context of decarbonizing freight transport, where the future development of new, renewable 
fuel technologies is inherently uncertain. 



To illustrate its capabilities, we apply STraM to a case study of the Norwegian freight transport sector. One important input here is data on the generalized transport costs of different transport modes and fuels. Because mode and fuel choice is a main driver for the model's output, we take extra care to use harmonized generalized transport cost data, by using and expanding on the cost model of \cite{Martin2023CarbonSectors}. 

The case study results show that the explicit modeling of multiple time periods leads to significantly different outcomes in terms of the resulting fuel mix and the associated investments made, compared to static model runs. Similarly, we demonstrate that explicitly considering uncertainty leads to a significant improvement in expected total system costs {compared to deterministic planning.} In terms of policy insights, we find that a carbon price is an effective instrument to achieve decarbonization of the freight transport system. Finally, the results show interesting regional differences in optimal infrastructure investments.

The remainder of this paper is organized as follows. 
In Section~\ref{sec:literature_review}, we discuss the literature on (national) freight transport models, with a specific focus on decarbonization, and identify relevant gaps. 
In Section~\ref{sec:model}, we introduce our general modeling approach in STraM, discuss the scope and main assumptions, and provide a mathematical formulation of our model. In Section~\ref{sec:case_study}, we apply the STraM framework on a case study of the Norwegian freight transport system and define all the required data input. In Section~\ref{sec:results}, we present results, and provide insights into the performance of our modeling framework. Section~\ref{sec:discussion} presents a discussion. Finally, Section~\ref{sec:conclusion} concludes the paper.

\section{Literature review} \label{sec:literature_review}

The literature contains various approaches to decision support models for decarbonizing national freight transport. 
For example, policy analyses, often targeting modal and/or fuel shift, have been performed using system dynamics models \citep{Nassar2023ATransportation,Ghisolfi2024DynamicsBrazil} and energy system models \citep{Rosenberg2023ModellingNorway, Hoehne2023ExploringMobility}. 
However, the representation of the underlying physical transport infrastructure in these approaches is often insufficient, making it difficult to analyze the efficacy of various specific infrastructure investments. An alternative modeling paradigm that recognizes the importance of modeling the transport network explicitly is the use of \textit{national freight transport models}, which we embrace.

In the remainder of this section, we first sketch the historical development of these models and their application to decarbonization of freight transport. Then, we explain how STraM builds upon and extends these existing approaches, to be able to provide insights into the effects of policy measures and infrastructure investments in a national freight transport system.



During early years, national freight transport models were typically set up  
as \textit{network design models}. Specific examples include STAN \citep{Guelat1990,Crainic1990StrategicSystem}, NODUS \citep{Jourquin1996TransportationEurope} and
TLSS \citep{Arnold2004ModellingSystem}. In this approach, the freight transport system is modeled "bottom-up" as a graph, and an optimization problem is formulated on this graph that finds infrastructure investments that minimize the sum of investment and operational costs of the entire system \citep{Crainic2021NetworkLogistics}. Here, the operational costs are computed by assigning transport demand between pairs of nodes to transport modes and routes in the graph. 
Examples of these models have been used in different countries, including Brazil, Sweden and Norway \citep{Guelat1990,DeJong2009DiscreteSweden}. 

Over the past decades, national freight transport models have shifted away from the simpler network design models. Instead, they have moved towards more elaborate, high-resolution models, which capture various tactical and operational planning elements \citep{Crainic1997}, and often resemble simulation models \citep{Crainic2018SimulationTaxonomy}.
In particular, many authors have focused on developing advanced logistics modules that accurately reflect the logistics decision-making processes in the freight transport system \citep{deJong2013RecentEurope, Archetti2022OptimizationSurvey}. 
For example, the Norwegian National Freight Transport Model \citep{AnneMadslien2015NasjonalModelen} uses a so-called aggregate-disaggregate-aggregate approach by \cite{Ben-Akiva2013TheSystem}, in which aggregate production and consumption levels are translated to disaggregated firm-level transportation demands. 
Many other national governments, including those of Canada, Germany, Finland, Italy, the Netherlands, Sweden, and the United Kingdom have developed similar models \citep{deJong2013RecentEurope}.

Most national freight transport models are based on the four-step procedure originally developed for passenger transport models \citep{deJong2004NationalDevelopment}. In this framework, the modeling task is divided into four steps. The first two steps, ``production/attraction'' and ``distribution'' consist of generating production and consumption levels at different locations in the network, and translating these to specific demands for transportation of goods (defined by an origin and destination location). The third step, ``modal split'', consists of dividing the demand over different modes of transport, such as road, sea, or rail. The final step, ``assignment'', assigns the demand to specific vehicles and routes in the network, given the selected mode. The STraM model, presented in this paper, focuses on the third and fourth step.

In recent years, national freight transport models have been applied to answer questions about decarbonization of freight transport. For example, \citet{deBok2020ExploringNetherlands} evaluate the effect of an emission-based truck charge on vehicle and shipment size choices, using the Dutch national freight transport model \mbox{BASGOED}. 
Furthermore, \citet{Wangsness2021Double-trackModel} consider the introduction of a double-track railway between Oslo and Gothenburg in the Norwegian Freight Transport model, and analyze its effect on carbon emissions. 
The existing national freight transport models are indeed very useful for providing insights into the effects of specific policy measures and infrastructure investments, as in these examples.

However, in the context of decarbonization of freight transport, the existing national freight transport models struggle to answer strategic questions involving, for example, the impact of uncertainty, the interaction between many investment options, or the development of the transport system over time. One issue is that these models need highly detailed input data, which is hard to obtain for future states of the system \citep{Meersman2019FreightFuture}. A more fundamental issue is that the existing models are static (one time period) and do not include long-term uncertainty, whereas strategic infrastructure decisions have effects that unfold over long periods of time and depend heavily on uncertain developments regarding, e.g., novel renewable fuels.


To deal with the inherent \textit{long-term} uncertainty, some authors perform what-if analyses on the existing static models \citep{Crainic1990ARail, Pinchasik2020CrossingNordics}.
However, the limitations of what-if analyses are well known \citep{Higle2003}. For example, an infrastructure investment that is suboptimal in every future scenario might actually be worthwhile when taking uncertainty into account, if it offers a form of flexibility to deal with many different scenarios. What-if analyses do not see the value of flexibility. 
This is problematic in strategic decision-making situations, in which infrastructure investments should be decided on today, but their effects take place over the coming decades and are inherently uncertain.

While research on long-term uncertainty in strategic freight transport models is limited, several works do address \emph{short-term} uncertainty. For example,  
\citet{Demir2016AUncertainty} deal with uncertainty in short-term transport demand and travel times in service network design modeling, and \citet{Yang2016PlanningUncertainty} consider uncertainty in both transportation cost and travel times for an intermodal freight transport planning problem. 

With STraM, we aim to fill the gaps identified above, by developing a strategic network design model with multiple time periods, a long planning horizon, and the consideration of long-term uncertainty. 

\section{Model} \label{sec:model}

In this section, we present STraM. We first provide an overview of its scope and the main assumptions, and then provide the mathematical formulation.

\subsection{Model scope and assumptions} \label{sec:scope}  

\begin{figure}[htbp!]
  \centering
  \includegraphics[width=.9\textwidth]{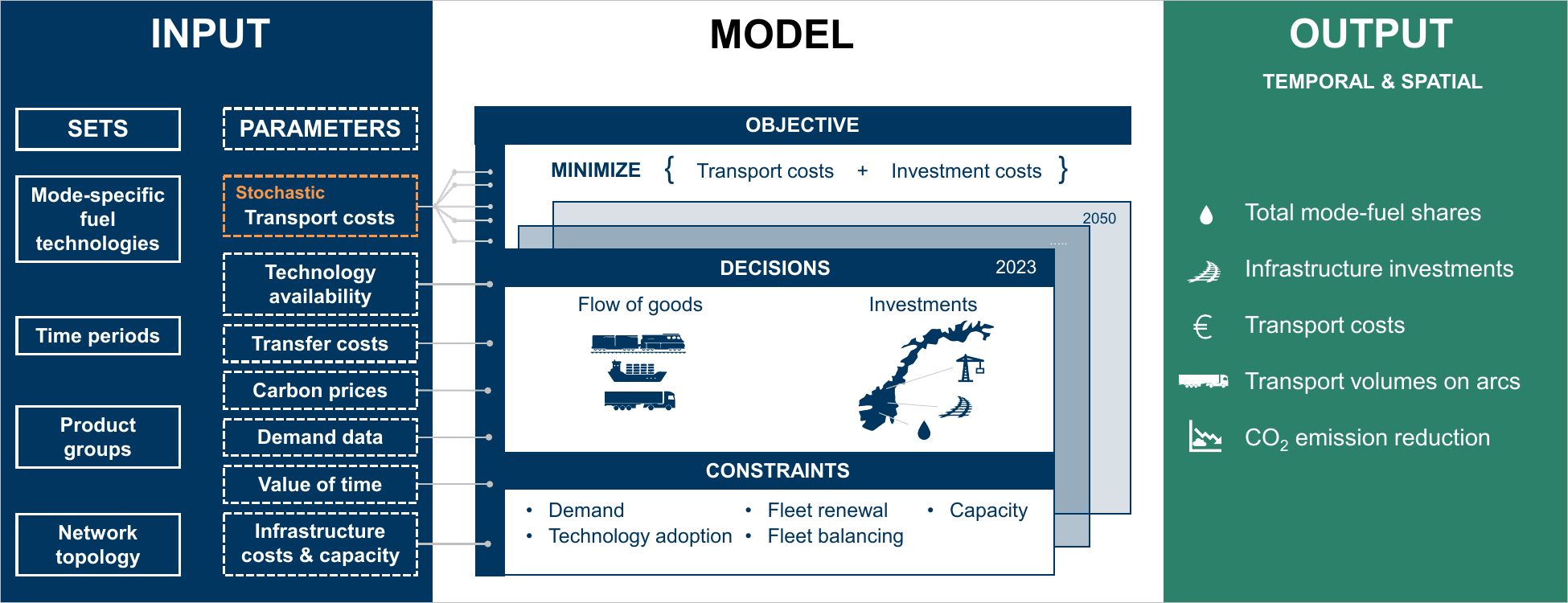}
  \caption{Conceptual overview of STraM}
  \label{fig:model_overview}
\end{figure}

STraM is a modeling framework for strategic national freight transport network design. The model takes the perspective of a social planner and decides on major infrastructure investments and flows of transport through a multimodal network. The objective is to minimize overall investment and transportation costs, including carbon fees. The output of the model is a collection of investment decisions in every time period and scenario, as well as a choice of transport mode, fuel, and route for each transportation demand. See Figure~\ref{fig:model_overview} for an overview of the model.

STraM focuses on \textit{line-haul} transport in a specific country. Line-haul refers to the transport of goods over long distances between two major points of interest, such as cities or harbors. It focuses on the main segment of the journey, excluding potential local or regional transport, and can include various transport modes, like road, rail, or sea. 
To achieve this, the country is represented by a graph, with each node corresponding to one of the country's regions. Demand for transport is aggregated by region and product group and represented as demand for transport between pairs of nodes. Transport within regions falls outside the scope of the model and is ignored. The available transport connections between regions are represented by the edges in the graph, with separate edges for different transport modes (e.g., sea, road, rail). On each edge, vehicles using different fuels (diesel, ammonia, battery-electric, etc.) can be used. This is modeled using a representative vehicle for each fuel and product group. 

There are limitations on how much each mode and fuel can be used. 
Freight transport on rail is restricted by limited capacity in both railways and terminals, while sea transport is only restricted by harbor capacity. Moreover, the use of specific fuels in each time period is restricted by the availability of necessary infrastructure (e.g., charging and filling infrastructure on road). These capacities can be expanded by means of infrastructure investments, which are the main strategic decisions in STraM.

We take a \textit{strategic} perspective by including a long planning horizon as well as uncertainty in the development of new fuel technologies. The system is modeled at different time periods from today until the planning horizon. Uncertainty is included through a set of scenarios for the random parameters in the model. We use a two-stage structure, meaning that the time periods are divided between a deterministic first-stage and a random second-stage. The first-stage decisions (e.g., infrastructure investments) should anticipate on the future uncertainty. 

\subsection{Mathematical formulation} \label{subsec:math_formulation}  
We now present the mathematical formulation of STraM. We first provide an overview of the sets, parameters and variables in the model. Then, we present the objective and constraints.

\subsubsection{Sets, parameters and variables}

\begin{figure}[htbp!]
     \centering
     \begin{subfigure}[b]{0.25\textwidth}
         \centering
         \includegraphics[width=\textwidth]{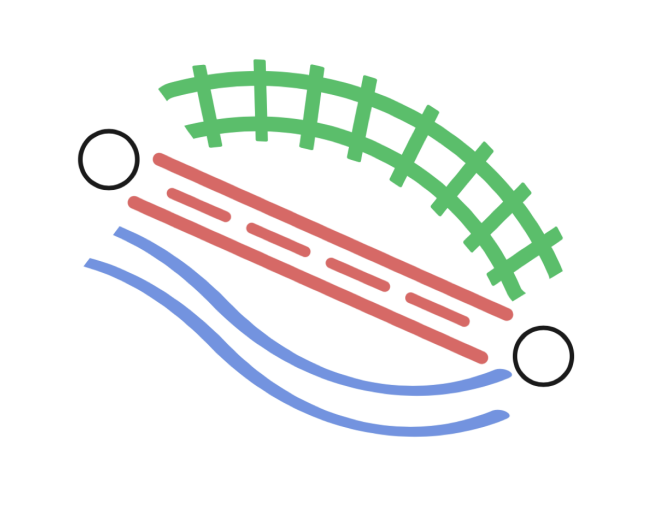}
         \caption{Three (undirected) edges.}
         \label{fig:three_edges}
     \end{subfigure}
     \hspace{10mm}
     \begin{subfigure}[b]{0.25\textwidth}
         \centering
         \includegraphics[width=\textwidth]{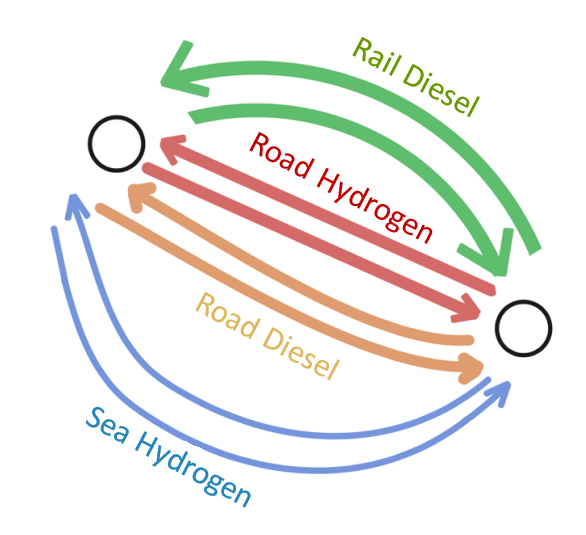}
         \caption{Eight (directed) arc-fuel combinations.}
         \label{fig:arcs}
     \end{subfigure}
        \caption{Example of three edges and eight arc-fuel combinations on the same pair of nodes.}
        \label{fig:edges_and_arcs}
\end{figure}

The underlying transportation network is modeled as a directed graph $\mathcal{G} = (\mathcal{N},\mathcal{A})$. Nodes $\mathcal{N}$ represent the major areas in a country, 
while directed arc $(i,j,m,r) \in \mathcal{A}$ represents the link from node $i\in\mathcal{N}$ to node $j\in\mathcal{N}$ for a transport mode $m\in\mathcal{M}$ and route alternative $r\in\mathcal{R}$. Route alternatives are used to model multiple ways to go from $i$ to $j$ using mode $m$, for example when there are parallel rail tracks going through different valleys. While flow of goods can be expressed in terms of the directed arcs, we also define the corresponding undirected edges $e \in \mathcal{E}$ to represent the underlying infrastructure. Along every arc $a = (i,j,m,r)$ on the path, different fuels $f \in \mathcal{F}_m$ can be used for transporting the goods. Figure~\ref{fig:edges_and_arcs} illustrates our use of arcs and edges.

Demand for transportation from origin $o \in \mathcal{N}$ to destination $d \in \mathcal{N}$ is differentiated by product group $p \in \mathcal{P}$. 
This demand can be fulfilled along paths $k \in \mathcal{K}$ in the network, which are defined as ordered sequences of directed arcs, i.e., $k=(a_1,...,a_n)$ for $a_1,...,a_n \in \mathcal{A}$. To restrict the feasible space of the model, only a limited set $\mathcal{K}$ of admissible paths is used. For each transport mode, we define a set of representative vehicles $\mathcal{V}_m$. We assume that transport of a given product group is always performed using the same representative vehicle type on each mode.

Several elements of the network are capacitated. First, some of the edges (e.g., those representing railroads) have a maximum capacity. Moreover, certain edges $e$ can be upgraded to allow the usage of a specific fuel $f$. These upgrade possibilities are collected in the set $\mathcal{U} \subseteq \mathcal{E}\times\mathcal{F}$ and most notably contain the electrification of certain rail edges. Additionally, within each node there can be different capacitated terminals that are used to load and unload goods at their origin and destination and to transfer goods between different modes along a path. 

The time dimension is modeled using a set of time periods $\mathcal{T} = \left\{ 0,1,...,T\right\}$, where period $0$ is the present year, and $T$ is the end of the planning horizon. Each period represents a number of years in the planning horizon. Uncertainty is modeled using a scenario tree that lives on this timeline, and is represented by the set $\mathcal{S}$ of scenarios. We assume a two-stage setting, with first-stage periods $\mathcal{T}^{1st}$ and second-stage periods $\mathcal{T}^{2nd}$. Section~\ref{sec:scen_gen} elaborates on the procedure to generate the scenario tree that describes the uncertainty.  

For each time period $t \in \mathcal{T} $ and scenario $s\in \mathcal{S}$ we define a range of decision variables. 
The annual flow (in tonnes) of a product type $p \in \mathcal{P}$ on a path $k \in \mathcal{K}$ is given by the continuous non-negative variable $h_{kpts}$. These path-flow variables are linked to the arc-flow variables $x_{apfts}$, representing the flow (in tonnes) of a product $p \in \mathcal{P}$ over arc $a \in \mathcal{A}$, where the index $f \in \mathcal{F}$ specifies which fuel type is being used. 
Due to geographical imbalances in transport demand, vehicles sometimes have to make empty trips. We model these using path-flow variables $\tilde{h}_{kvts}$ and arc-flow variables $b_{afvts}$.
From these flow variables, we can derive the total transport amount (in tonne-km) of a mode-fuel combination $(m,f)$ throughout the entire network, which we denote by $q_{mfts}$. 

The strategic decisions in the model are investments in transport infrastructure. Investment in terminal capacities are represented by the binary variable $\nu_{ncmts}$, representing a one-time capacity expansion in node $i$ of terminal type $c$ for mode $m$. Investment in edge capacities are represented by the binary variables $\varepsilon_{ets}$. The charging/fueling capacity of a fuel $f$ on edge $e$ can be increased by means of the continuous variable $y_{efts}$. Finally, infrastructure on an edge can be upgraded to allow for the use of a new fuel $f$ using the binary variable $\upsilon_{efts}$. For all these investment decisions, we assume that the investment is finished after a predefined lead time.

See Tables~\ref{tab:sets}-\ref{tab:variables} for a complete overview of all sets, parameters and variables in the model.

\begin{table}[h!]
\footnotesize 
\centering
\caption{Sets and indices}
\label{tab:sets}
\begin{tabular}{@{}ll@{}}
\toprule
    $\mathcal{A} $      & Set of directed arcs, indexed by $a$, where $a=(i,j,m,r)$. \\
                        &is defined from node $i$ to node $j$ with mode $m$ on route $r$. \\
    $\mathcal{A}_{e} \subseteq\mathcal{A} $ & Arcs $a$ on edge $e$. \\
    $\mathcal{A}_{m} \subseteq\mathcal{A} $ & Arcs $a$ using mode $m$. \\
    $\mathcal{A}^{\text{in}}_{nm} \subseteq\mathcal{A} $ & Arcs $a=(i,n,m,r)$ entering node $n$, using mode $m$. \\
    $\mathcal{A}^{\text{out}}_{nm} \subseteq\mathcal{A} $ & Arcs $a=(n,j,m,r)$ leaving node $n$, using mode $m$. \\
    $\mathcal{C}_m $ & Set of possible terminal classes for mode $m$, indexed by $c$.\\
    $\mathcal{E} $ & Set of undirected edges defined as $\mathcal{E}=\text{undir}(\mathcal{A})$, indexed by $e$. \\
    $\mathcal{E}_m \subseteq\mathcal{E} $ & Set of edges $e$ for a specific mode $m$.\\
    $\mathcal{F}     $ & Set of fuels, indexed by $f$.\\
    $\mathcal{F}_{a} \subseteq \mathcal{F} $ & Set of allowed fuels on arc $a$.\\
    $\mathcal{F}_{m} \subseteq \mathcal{F} $ & Set of allowed fuels for a mode $m$.\\
    $\mathcal{F}_{m}^{\text{\text{new}}} \subseteq \mathcal{F} $ & Set of new fuels that are allowed on mode $m$.\\
    $\mathcal{K}     $ & Set of paths, indexed by $k$. \\
    $\mathcal{K}_a \subseteq \mathcal{K}    $ & Set of paths that include arc $a$. \\
    $\mathcal{K}_{od} \subseteq \mathcal{K}   $ & Set of paths that lead from origin $o$ to destination $d$. \\
    $\mathcal{K}_{im}^{\text{cap}} \subseteq \mathcal{K}   $ & Set of all paths that use terminal capacity in node $i$ for mode $m$. \\
    $\mathcal{K}^{\text{uni}} \subseteq \mathcal{K}   $ & Set of all unimodal paths. \\
    $\mathcal{K}_{m}^{\text{uni}} \subseteq \mathcal{K}   $ & Set of all unimodal paths using mode $m$. \\
    $\mathcal{M}     $ & Set of modes, indexed by $m$. We have $\mathcal{M}=\left\{Rail,Road,Sea\right\}$. \\
    $\mathcal{N}     $ & Set of nodes, indexed by $n, i, j, o$, or $d$.\\
    $\mathcal{N}_m\subseteq\mathcal{N} $ & Set of all nodes with the possibility to invest in terminals for mode $m$. \\ 
    $\mathcal{P}  $ & Set of product groups, indexed by $p$. \\
    $\mathcal{P}_{c} $ & Set of product groups that can be processed at terminal class $c$. \\
    $\mathcal{R}     $ & Set of route alternatives, indexed by $r$. \\
    $\mathcal{S} $ & Set of scenarios, indexed by $s$.\\
    $\mathcal{S}^t_s \subseteq \mathcal{S} $ & Set of scenario's that share the same history as scenario $s$, \\ 
                                            &up to, and including, time period $t$. \\
    $\mathcal{T} $ & Set of time periods, indexed by $t$. \\
    $\mathcal{T}^{1st} \subseteq \mathcal{T}$ & Set of first-stage periods, indexed by $t$. \\
    $\mathcal{T}^{2nd} \subseteq \mathcal{T}$ & Set of second-stage periods, indexed by $t$. \\
    $\mathcal{T}_t^g $ & Set of time periods for which, when investing in $g \in \left\{ \text{edge}, \text{node}, \right.$ \\
    &$\left.\text{upgr},\text{charge}\right\}$, the investment is finished by time period $t$. \\
    $\mathcal{U} \subseteq \mathcal{E}\times\mathcal{F} $ & Set of possible upgrades in terms of edge-fuel combinations. \\
    $\mathcal{V}_m$ & Set of vehicle types on mode $m$.  \\
    $\mathcal{V}_a$ & Set of vehicle types on mode $m$ used in arc $a = (i,j,m,r)$.  \\
    $\mathcal{Y}$ & Set of all years within the planning horizon.\\
    \bottomrule
    \end{tabular}
\end{table}
\begin{table}[h!]
    \footnotesize
    \centering
    \caption{Parameters}
    \label{tab:parameters}
    \begin{tabular}{@{}ll@{}}
        \toprule
        $A_{mft} \in \left[0,1 \right]$            & Theoretical maximum technology adoption level of a mode-fuel combination \\
                                    & in time period $t$ based on the associated Bass diffusion model.  \\
        ${C^{}_{afpts}}        $     & Generalized cost of transporting one tonne of product $p$, using arc $a$ and \\
                                &fuel $f$ in time period $t$ and scenario $s$ (EURO/tonne). \\
        ${\tilde{C}^{}_{afvts}}        $     & Generalized cost of transporting one tonne of empty capacity of vehicle $v$, \\
                                    &using arc $a$ and fuel $f$ in time period $t$ and scenario $s$ (EURO/tonne). \\
        ${C^{\text{transf}}_{kp}}     $  & Total transfer costs on path $k$ for product $p$ (EURO/tonne). \\
        ${C^{\text{charge}}_{af}}   $      & Unit cost of increasing charging/filling capacity on arc $a$ for fuel $f$ \\ & (EURO/tonne). \\
        ${C^{\text{edge}}_{e}}     $       & Investment cost of increasing the capacity of edge $e$ (EURO). \\
        ${C^{\text{node}}_{ncm}}     $     & Investment cost of increasing the capacity at node $n$ for terminal class $c$ and \\
                                        &mode $m$ (EURO).\\
        ${C^{\text{upg}}_{e}}     $        & Investment cost of upgrading edge $e$ (EURO).\\
        $D_{opt} $                  &Demand for transport of product $p$ from origin $o$ to destination $d$ in time \\
        & period $t$ (tonnes).\\
        $L_{a}      $               & Length of an arc $a$ (km).\\
        $M_{}             $        & Sufficiently large number (big M). \\
        $N_{m}             $   & Lifespan of the representative vehicle for mode $m$ (years).\\
        $P_{s} \in \left[0,1 \right] $                & Probability assigned to scenario $s$. \\
        ${\overline{Q}^{\text{charge}}_{ef}} $ &   Initial charging/filling capacity of edge $e$ for fuel $f$ (tonnes). \\ 
        ${\overline{Q}^{\text{edge}}_{e}} $ & Initial capacity of edge $e$ (tonnes).\\
        ${Q^{\text{edge}}_{e}}     $       & Capacity increase from investing in infrastructure on edge $e$ (tonnes).\\
        ${\overline{Q}^{\text{node}}_{ncm}}$ & Initial capacity in node $n$ for terminal class $c$ for mode $m$  (tonnes). \\
        ${Q^{\text{node}}_{ncm}} $         & Capacity increase from investing in node $n$, terminal class $c$ and mode $m$ \\& (tonnes).\\
            $U_{mf} \in [0,1]$ & Potential adoption share of mode $m$ and fuel $f$.\\
        $Y_t $                      & Years passed at the beginning of time period $t$ (years).\\
        $\alpha_{mf\tau s}$ & Coefficient of innovation in the Bass            diffusion model   for mode-fuel \\
        & combination $(m,f)$ in year $\tau$ in scenario $s$. \\
        $\beta_{mf\tau s}$ & Coefficient of imitation in the Bass diffusion model for mode-fuel   \\
        & combination $(m,f)$ in year $\tau$ in scenario $s$. \\
        $\delta \in \left[0,1 \right]          $        & Yearly discount factor. \\
        $\rho_{mt} \in \left[0,1 \right]          $        & Maximum allowed relative decrease in transport work of mode $m$ \\
        &compared to time period $t-1$. \\
        \bottomrule
    \end{tabular}
\end{table}

\begin{table}[h!]
\footnotesize
\centering
\caption{Variables}
\label{tab:variables}
\begin{tabular}{@{}ll@{}}
\toprule
\multicolumn{2}{l}{\textit{Continuous variables}}     \\
    $b_{afvts}$         & Flow of balancing movements over arc $a$ using fuel $f$ and vehicle type $v$ \\ & in time period $t$ in scenario $s$ (tonnes of carrying capacity). \\
    $h_{kpts}$          & Flow of product $p$ on path $k$ in time period $t$ in scenario $s$ (tonnes). \\
    $\tilde{h}_{kvts}$    & Flow of balancing movements for vehicle type $v$ on path $k$ in time period $t$ \\
                        &in scenario $s$ (tonnes of carrying capacity). \\
    $q_{mfts}$          & Total transport amount using mode-fuel combination $(m,f)$ in  \\
                        & period $t$ in scenario $s$ (tonne-km).  \\
    $\tilde{q}_{mf\tau s}$          & Total transport amount using mode-fuel combination $(m,f)$ in  \\
                        & year $\tau$ in scenario $s$ (tonne-km).  \\
    $\tilde{q}^{\text{tot}}_{m\tau s}$  & Total transport amount on mode $m$ in year $\tau$ in scenario $s$ (tonne-km). \\
    $q^-_{mfts}$        & Total decrease in transport amount using mode-fuel combination $(m,f)$\\
                        & from period $t-1$ to period $t$ in scenario $s$ (tonne-km). \\
    $ x_{afpts}$        & Flow of product $p$ over arc $a$ using fuel $f$ in time period $t$ in scenario $s$ (tonnes).\\
    $y^{}_{efts}$       & Increase in charging/filling capacity on edge $e$ for fuel $f$ in time period $t$ in \\
                        & scenario $s$ (tonnes).
    \\[3mm]
\multicolumn{2}{l}{\textit{Binary variables}} \\
    ${\upsilon_{efts}}$     & $1$ if edge $e$ is upgraded in time period $t$ in scenario $s$ to allow fuel $f$; $0$ otherwise. \\
    ${\varepsilon_{ets} }$        & $1$ if the capacity of edge $e$ is expanded in time period $t$ in scenario $s$; $0$ otherwise. \\
    ${\nu_{ncmts}} $      & $1$ if in node $n$ the capacity of terminal class $c$ for mode $m$ is expanded in time  \\ 
                        & period $t$ in scenario $s$; $0$ otherwise. \\* \bottomrule
\end{tabular}
\end{table}

\subsubsection{Objective function}

The main objective is to minimize the total (generalized) costs of transport and investment in infrastructure. We first discuss how the total costs are computed for each scenario and then how the scenarios are aggregated in a way that reflects risk-averse preferences.

The total discounted costs $f(\xi,\mathbf{x})$ are a function of the uncertain parameters summarized as $\xi$ and the first-stage decision variables summarized as $\mathbf{x}$. For scenario $s \in \mathcal{S}$, it is defined as:
\begin{alignat}{2}
        f(\xi^s,\mathbf{x}):=&   \sum_{t\in\mathcal{T}} \left[ \sum_{l=Y_t}^{Y_{t+1}-1} \delta^l\left(  \sum_{a \in \mathcal{A}} \sum_{f\in\mathcal{F}_a} 
        \left\{
        \sum_{p\in\mathcal{P}}   C_{afpts} x_{afpts}  +  
        \sum_{v\in\mathcal{V}_a} \tilde{C}_{afvts} b_{afvts}  \right\}  \right. \right.
          &&    \notag \\%
        & 
        + \left. \sum_{p\in\mathcal{P}} \sum_{k \in \mathcal{K}} C^{\text{transf}}_{kp}h_{kpts}   \right)
        +	\delta^{Y_t}\left(
        \sum_{m \in \left\{\text{sea}, \text{rail} \right\}} \sum_{n \in \mathcal{N}_{m} } \sum_{c \in \mathcal{C}_{m}} C^{\text{node}}_{ncm} \nu_{ncmts} \right.
         && \label{block:obj1} \\ %
        &   
        + \left.\left.   \sum_{e \in \mathcal{E}_{Rail}}C^{\text{edge}}_{e} \varepsilon_{ets} + \sum_{(e,f) \in\mathcal{U}} C^{\text{upg}}_{ef} \upsilon_{efts}+  
        \sum_{e \in \mathcal{E}_{\text{road}}} \sum_{f\in\mathcal{F}_m} C^{\text{charge}}_{ef}y^{}_{efts}    \right)
        \right]. && \notag
\end{alignat}
The cost function consists of a discounted sum of transport costs (within the first pair of round brackets) and investment costs (within the second pair of round brackets). The annual transport costs consist of: (i) the generalized cost of transporting the goods over the arcs in the network (including balancing trips); and (ii) the cost of transferring goods from one transport mode to another along the routes. The investment costs consist of (i) the cost of expanding terminal capacities, (ii) the cost of expanding edge capacities, (iii) the cost of upgrading edges to allow more fuel types, and (iv) the cost of building charging/filling stations along edges. For each year in a period $t$, the yearly transport costs, including carbon fees, are discounted and added separately. In contrast, investment costs are only added once.

In strategic planning situations, it is common for a social planner (government) to have a preference for solutions that are cost-efficient, but also robust across different future scenarios. To model such risk-averse preferences regarding uncertainty, we use a so-called mean-CVaR objective function. That is, the objective function is a weighted average of the expected value and the conditional value at risk (CVaR) of the total discounted costs $f(\xi,\mathbf{x})$: 
\begin{align}
    \min_{\mathbf{x}} \left\{ (1 - \lambda) \mathbb{E}\big[ f(\xi,\mathbf{x}) \big] + \lambda \text{CVaR}_\gamma\big(f(\xi,\mathbf{x})\big) \right\}, \label{obj:main}
\end{align}
where $\lambda \in [0,1]$ represents the relative weight of CVaR compared to the expectation. This formulation strikes a balance between good performance on average and avoiding risk, which is captured by CVaR: the expected value of the $\gamma$-tail of the distribution of the total costs \citep{Rockafellar2002ConditionalDistributions}. 

The choice of CVaR as our risk measure has two main advantages. First, it is coherent \citep{Artzner1999CoherentRisk}, meaning that it satisfies a number of properties that are desirable to accurately reflect risk-averse preferences. The second advantage is that CVaR can be computed using a minimization problem:
\begin{align}
    \text{CVaR}_\gamma(Z) := \min_{u \in \R} \left\{  u + (1 - \gamma)^{-1} \mathbb{E}\big[ (Z - u)^+ \big] \right\}.
\end{align}
This representation allows us to recast the risk-averse problem as a standard risk-neutral stochastic program 
\citep{Rockafellar2002ConditionalDistributions} with an additional first-stage decision variable $u$. This is indeed how we solve our mean-CVaR model.

\subsubsection{Constraints}

The \textit{demand}, \textit{arc-path} and \textit{fleet balancing} constraints are given by:
\begin{alignat}{2}
    & \sum_{k \in \mathcal{K}_{od}} h_{kpts} = D_{odpt},  \qquad&  o,d \in \mathcal{N}, \: p \in \mathcal{P}, \: t \in \mathcal{T}, \: s \in \mathcal{S},	&
    \label{block:demand1} \\
    & \sum_{f \in \mathcal{F}_a} x_{afpts} = \sum_{k \in \mathcal{K}_a} h_{kpts},\quad& a \in \mathcal{A}, \: p \in \mathcal{P}, \: t \in \mathcal{T}, \: s \in \mathcal{S}, &
    \label{block:path_arc1} \\
    & \sum_{f \in \mathcal{F}_a} b_{afvts} = \sum_{k \in \mathcal{K}_a \cap \mathcal{K}^{\text{uni}}_m} \tilde{h}_{kvts}, \quad \quad \quad \quad \quad \quad \quad \quad   & a \in \mathcal{A}, \: v \in \mathcal{V}_a, \: t \in \mathcal{T}, \: s \in \mathcal{S},
    \label{block:path_arc_balancing} \\
    & \mathrlap{   \sum_{a \in \mathcal{A}_{nm}^{\text{in}}} \left( b_{afvts} + \sum_{p \in \mathcal{P}_v} x_{afpts} \right) = \sum_{a \in \mathcal{A}_{nm}^{\text{out}}} \left( b_{afvts} + \sum_{p \in \mathcal{P}_v} x_{afpts} \right) }& \nonumber\\
    &  & \mathllap{n \in \mathcal{N}, \: m \in \mathcal{M}, \: f \in \mathcal{F}_m, \: v \in \mathcal{V}_m, \: t \in \mathcal{T}, \: s \in \mathcal{S}.}
    \label{block:fleet_balance} 
\end{alignat}
Here, constraints~\eqref{block:demand1} ensure that the \textit{demand} for transport between all origins and destinations is satisfied. Constraints~\eqref{block:path_arc1}~and~\eqref{block:path_arc_balancing} describe the \textit{arc-path} relation for the freight flow variables and for the vehicle balancing variables, respectively. 
The \textit{fleet balancing} 
constraints~\eqref{block:fleet_balance} ensure that 
every vehicle that leaves a node eventually comes back. This constraint sometimes induces trips with empty vehicles, 
a feature that is present in \citet{Crainic1990ARail}, but has been neglected in some more recent models.

\textit{Capacity} constraints 
limit the flow of goods on nodes and edges:
\begin{subequations}
\label{block:capacity_constraints}
\begin{alignat}{2}
        &   \mathrlap{\sum_{f \in \mathcal{F}_{a}} \left( \sum_{p\in\mathcal{P}}  x_{afpts} +  \sum_{v \in\mathcal{V}_a}  b_{afvts} \right) \leq \frac{1}{2}\left(\overline{Q}^{\text{edge}}_{e} + Q^{\text{edge}}_{e}\sum_{t^{'} \in \mathcal{T}_{et}^{\text{edge}}} \varepsilon_{e t^{'} s}\right),}  \quad& \notag \\
        & \quad& e \in\mathcal{E}_{\text{rail}}, \: a\in\mathcal{A}_{e}, \: t \in \mathcal{T}, \: s \in \mathcal{S},
        \label{block:edge_capacity1} \\
         &	\sum_{t \in\mathcal{T}} \varepsilon_{e t s} \leq 1 ,	
         \quad&	e \in \mathcal{E}_{\text{rail}}, \: s \in \mathcal{S},
         \label{block:edge_capacity2} \\
        & \sum_{p\in\mathcal{P}_c}\sum_{k\in\mathcal{K}^{\text{cap}}_{im}} h_{kpts}  \leq \overline{Q}^{\text{node}}_{ncm} + Q^{\text{node}}_{ncm}\sum_{t^{'} \in \mathcal{T}_{ncmt}^{\text{node}}} \nu_{ncmt^{'} s},   \quad&   \notag \\
        &   \quad& 	\mathllap{m \in \left\{\text{sea}, \text{rail}\right\}, \: n \in \mathcal{N}_{m}, \: c\in \mathcal{C}_{m},  \ t \in \mathcal{T}, \: s \in \mathcal{S}},
        \label{block:node_capacity1}\\
        &	\sum_{t \in\mathcal{T}} \nu_{ncmts} \leq 1 , 	
        \quad&	 \mathllap{m \in \left\{ \text{sea}, \text{rail} \right\}, \: n \in \mathcal{N}_{m}, \: c \in \mathcal{C}_{m}, \: s \in \mathcal{S}},
        \label{block:node_capacity2} \\
        &   \mathrlap{\sum_{a\in\mathcal{A}_{e}} \left( \sum_{p\in\mathcal{P}} x_{afpts} +
        \sum_{v \in\mathcal{V}_a}  b_{afvts} \right) \leq \overline{Q}^{\text{charge}}_{ef}+\sum_{t^{'} \in \mathcal{T}_{eft}^{\text{charge}}} y^{}_{eft^{'} s} ,}  \quad& \notag\\
        & \quad&	e \in \mathcal{E}_{\text{road}}, \: f\in \mathcal{F}_e, \: t \in \mathcal{T}, \: s \in \mathcal{S}, 
        \label{block:charge1} \\
        & \sum_{p\in\mathcal{P}}\sum_{a\in\mathcal{A}_{e}}  x_{afpts} \leq M_{} \sum_{t^{'} \in \mathcal{T}_{eft}^{\text{upg}}} \upsilon_{ef t^{'} s}, 
        \quad& (e,f) \in \mathcal{U}, \: t \in \mathcal{T}, \: s \in \mathcal{S}.   \label{block:upgrade1}
\end{alignat}
\end{subequations}
Constraints~\eqref{block:edge_capacity1}~and~\eqref{block:edge_capacity2} limit the flow of goods on arcs by the corresponding edge capacity, which is split equally between the arcs going in opposite directions.  
Similarly, constraints \eqref{block:node_capacity1} and \eqref{block:node_capacity2} limit the amount of goods transferred from one mode to another in a given node by the corresponding terminal capacity.
Constraints~\eqref{block:charge1} model the fact that for certain fuels, transport can only take place on edges that have necessary charging/filling infrastructure installed. Finally,~\eqref{block:upgrade1} models that some fuels can only be used on edges that have been upgraded.
 
\textit{Fleet renewal} constraints  
prevent the fleet mix from changing too suddenly, to model the fact that significant inertia exists in transport fleets: 
\begin{subequations}
\label{block:fleet_renewal}
\begin{alignat}{2}
& q_{mfts} = \sum_{p\in\mathcal{P}}\sum_{a\in \mathcal{A}_{m}}  L_a x_{afpts}, 
\hspace{3cm}\quad& {m \in \mathcal{M}, \: f\in\mathcal{F}_m, \: t \in \mathcal{T}, \: {s\in \mathcal{S}}},\label{block:transport_amount} \\
& q^-_{mfts} \geq q_{mf(t-1)s} - q_{mfts}, \quad & {m \in \mathcal{M}, \: f \in \mathcal{F}_m, \: t \in \mathcal{T} \setminus \{0\}, \: {s\in \mathcal{S}}}, \label{block:fleet_renewal_pospart} \\
& \sum_{f \in \mathcal{F}_m} q^-_{mfts} \leq \frac{Y_t - Y_{t-1}}{N_{m}} \tilde{q}^{\text{tot}}_{m(t-1)s}, \quad & {m \in \mathcal{M}, \: t \in \mathcal{T} \setminus \{0\}, \:  {s\in \mathcal{S}}}, \label{block:fleet_renewal_main}\\
&  \sum_{f \in \mathcal{F}_m} q_{mfts} \geq  (1-\rho_{mt}) \  \sum_{f \in \mathcal{F}_m} q_{mf,t-1,s},  \quad& {m \in \mathcal{M}, \: t \in \mathcal{T}\setminus \{0\}, \: {s\in \mathcal{S}}.}\label{block:limit_modal_decrease} 
\end{alignat}
\end{subequations}
Constraints~\eqref{block:transport_amount} define the total transport amount (in tonne-km) of a mode-fuel combination, which is a proxy of the number of vehicles used. 
Constraints~\eqref{block:fleet_renewal_pospart}, in combination with the non-negativity constraints on $q_{mfts}^-$, define the latter variable as the decrease in the total transport amount of mode-fuel $(m,f)$ from period $t-1$ to period $t$. These variables are used in constraints~\eqref{block:fleet_renewal_main} to limit the total decrease in the fleet of vehicles on each mode $m$. 
Constraints~\eqref{block:limit_modal_decrease} limit decreases in modal transport work over time, capturing that existing supply chains do not change too rapidly.

\textit{Technology adoption} constraints~\eqref{block:maturity_constraints} restrict the (rate of) adoption of new fuel technologies:
\begin{subequations}
\label{block:maturity_constraints}
\begin{alignat}{2}
& \tilde{q}_{m f, Y_t, s}  = q_{m f t s}, \quad & m \in \mathcal{M}, \: f \in \mathcal{F}_m, \: t \in \mathcal{T}, \: s \in \mathcal{S}, \label{block:q_tilde_def} \\
& \tilde{q}^{\text{tot}}_{m\tau s} = \sum_{f \in \mathcal{F}_m} \tilde{q}_{mf\tau s}, \quad& m \in \mathcal{M}, \: \tau \in \mathcal{Y}, \: s \in \mathcal{S}, \label{block:q_tot_tau} \\
& \tilde{q}_{mf\tau s} \leq  A_{mf\tau} \ \tilde{q}^{\text{tot}}_{m\tau s}, \quad&	m \in \mathcal{M}, \: f \in \mathcal{F}^{\text{new}}_m, \: \tau \in \mathcal{Y}, \: s \in \mathcal{S}, &		\label{block:maturitylimit1} \\
& \mathrlap{\tilde{q}_{mf\tau s} - \tilde{q}_{mf,\tau -1,s} \leq  \alpha_{mf,\tau-1,s} \ U_{mf} \ \tilde{q}^{\text{tot}}_{m,\tau-1,s} + \beta_{mf,\tau-1,s} \ \tilde{q}_{mf,\tau-1,s},} \nonumber \\
&\hspace{220pt}\quad& m \in \mathcal{M},  \ f \in \mathcal{F}^{\text{new}}_m,  \ \tau \in \mathcal{Y}\setminus\{0\},  \ s \in \mathcal{S}. \label{block:maturitylimit2}
\end{alignat}
\end{subequations}
Adoption of new fuel technologies (i.e., mode-fuel pairs $(m,f)$) is expressed in terms of the variables $\tilde{q}_{mf \tau s}$, which are yearly counterparts to the variables $q_{mfts}$ and defined in constraints~\eqref{block:q_tilde_def}, and constraints~\eqref{block:q_tot_tau} defining the total transport amount on mode $m$ for year $\tau$. The remaining constraints are based on the diffusion model for technology adoption by \cite{Bass1969ADurables}. The original Bass model is governed by the Bass equation $\frac{\frac{d}{dt}F(t)}{1-F(t)} = \alpha(t) + \beta(t) F(t)$, where $F(t)$ is the extent to which a technology has been adopted, as a fraction of its potential $U$, $\alpha(t)$ is called the coefficient of innovation, describing the base rate of adoption, and $\beta(t)$ is the coefficient of imitation, describing how the adoption rate increases at higher levels of adoption. 
We use the adoption level $A_{mf\tau}$ of a new technology $(m,f)$ according to the Bass model as an upper bound in constraints~\eqref{block:maturitylimit1}. Moreover, we extract constraints~\eqref{block:maturitylimit2} on the \textit{rate} of adoption as follows. We linearize the Bass equation by (1) ignoring the fraction $\frac{1}{1-F(t)}$ on the left-hand side (which is close to one at low adoption levels), (2) writing it in terms of the absolute adoption level rather than the fraction, and (3) discretizing the equation. We then include the equation as an inequality, yielding an upper bound on the adoption rate of a new technology.

\textit{Non-anticipativity} constraints~\eqref{eq:nac} are required to model the fact that the scenario is learned after the first-stage, i.e., the first-stage decisions should be the same in every scenario:
\begin{align}
    \mathbf{x}_{ts} - \mathbf{x}_{ts'} = 0, \quad t \in \mathcal{T}^{1st}, \: s, s' \in \mathcal{S}, \label{eq:nac}
\end{align}
where $\mathbf{x}_{ts}$ is defined as the vector consisting of all decision variables that are indexed by period $t$ and scenario $s$.

Finally, the \textit{domains} of the variables are given by:
\begin{subequations}
\label{block:domains}
\begin{alignat}{2}
        & b_{afvts} \geq 0, \quad& \mathllap{a \in \mathcal{A}, \: f \in \mathcal{F}_a, \: v \in \mathcal{V}_a, \: t \in \mathcal{T}, \: s \in \mathcal{S}}, &
        \label{block:domain_b}	\\
        &   h_{kpts} \geq 0, \quad& k \in \mathcal{K}, \: p \in \mathcal{P}, \: t \in \mathcal{T}, \: s \in \mathcal{S}, &
        \label{block:domain_h}	\\
        &   \tilde{h}_{kvts} \geq 0, \quad&  k \in \mathcal{K}^{\text{uni}}, \: v \in \mathcal{V}, \: t \in \mathcal{T}, \: s \in \mathcal{S}, &
        \label{block:domain_hbar}	\\
        &   q_{mfts} \geq 0, \quad& m \in \mathcal{M}, \: f \in \mathcal{F}_m, \: t \in \mathcal{T}, \: s \in \mathcal{S},
        \label{block:domain_qmf}	\\
        &   q^-_{mfts}\geq 0, \quad& m \in \mathcal{M}, \: f \in \mathcal{F}_m, \: t \in \mathcal{T} \setminus \{0\}, \: s \in \mathcal{S},
        \label{block:domain_qminusmf}	\\
        & \tilde{q}_{mf\tau s} \geq 0, \quad & m \in \mathcal{M}, \: f \in \mathcal{F}_m, \: \tau \in \mathcal{Y}, \: s \in \mathcal{S}, \label{block:domain_qtilde} \\
        & \tilde{q}^{\text{tot}}_{m\tau s} \geq 0, \quad& m \in \mathcal{M}, \: \tau \in \mathcal{Y}, \: s \in \mathcal{S},
        \label{block:domain_qtilde_tot}	\\
        & x_{afpts} \geq 0, \quad& \mathllap{a \in \mathcal{A}, \: f \in \mathcal{F}_a,  \ p \in \mathcal{P}, \: t \in \mathcal{T}, \: s \in \mathcal{S}}, &
        \label{block:domain_x}	\\
        &   y^{}_{efts}\geq 0, \quad& e \in \mathcal{E}_{\text{road}}, \: f\in\mathcal{F}_e, \: t \in \mathcal{T}, \: s \in \mathcal{S}, &
        \label{block:domain_y} \\
        &   z_{ts}\geq 0, \quad& t \in \mathcal{T}, \: s \in \mathcal{S}, &
        \label{block:domain_u}	\\
        &   \upsilon_{efts}\in\left\{0,1\right\}, \quad& (e,f) \in \mathcal{U}, \: t \in \mathcal{T}, \: s \in \mathcal{S}, &
        \label{block:domain_z}	\\
        &   \varepsilon_{ets}\in\left\{0,1\right\}, \quad& e \in \mathcal{E}_{\text{rail}}, \: t \in \mathcal{T}, \: s \in \mathcal{S}, &
        \label{block:domain_v}	\\
        &   \nu_{ncmts}\in\left\{0,1\right\},  \quad&  m \in \left\{\text{sea}, \text{rail}\right\}, \: n \in \mathcal{N}_m, \: c \in \mathcal{C}_m, \: t \in \mathcal{T}, \: s \in \mathcal{S}. &
        \label{block:domain_w}	
\end{alignat}
\end{subequations}

\section{Case study: Norway} \label{sec:case_study}

We apply STraM in a case study of the Norwegian freight transport system, and refer to this implementation as STraM-Norway. This case study has a dual purpose: to illustrate the capabilities and advantages of STraM and to gain insight in the decarbonization of the Norwegian freight transport system. As a time resolution, we use the years $2023$, $2028$, $2034$, $2040$, and $2050$, and assume the system between any subsequent pair of years operates as in the first of the two.

\subsection{Network topology} \label{sec:netwerk_topo}

\begin{figure}%
    \centering
    \subfloat[\centering Demand as sum of import and export activities per regional zones.]{
    \includegraphics[height=0.2\paperheight]{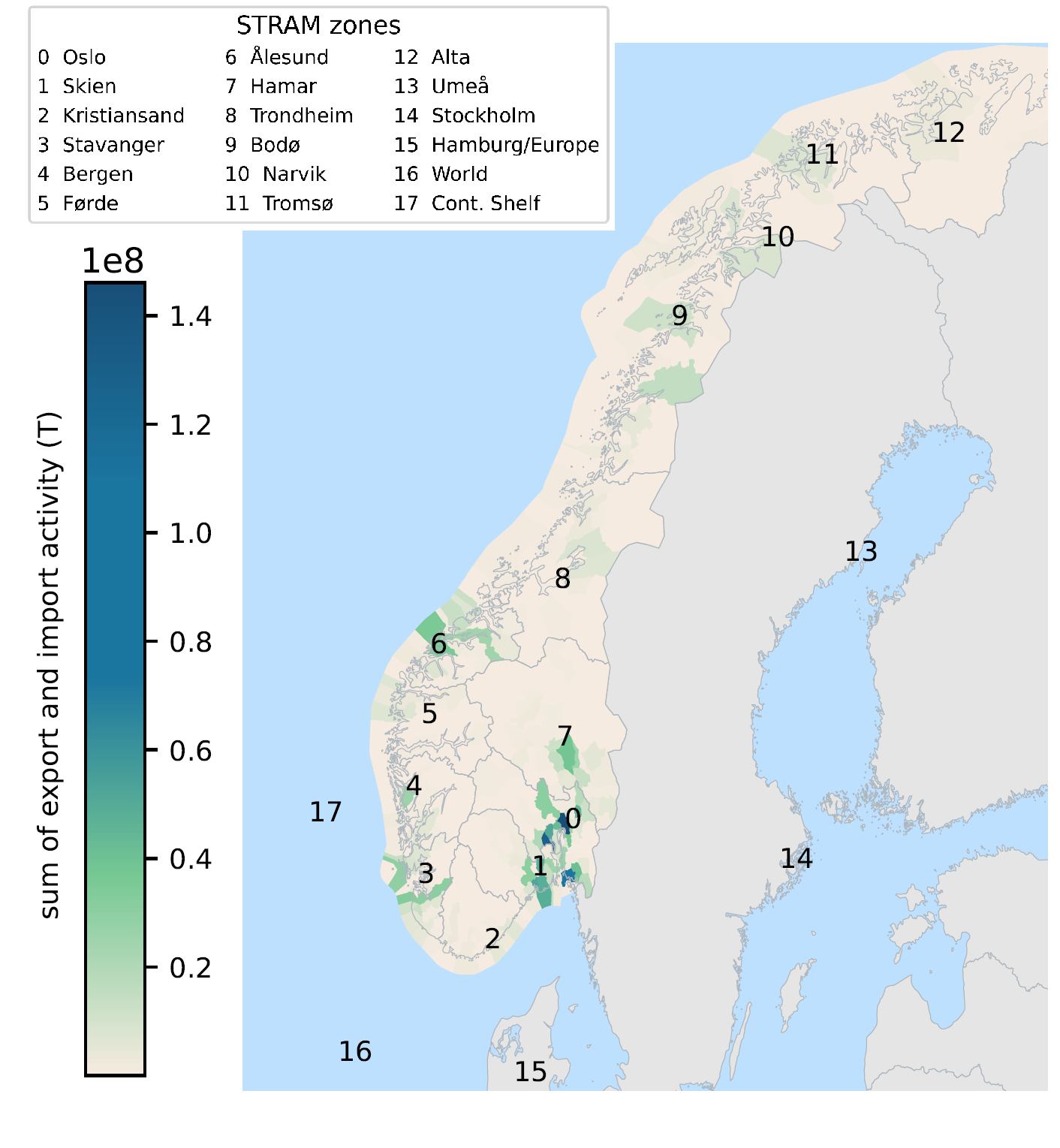}
    \label{fig:zones_demand}
    }%
    \subfloat[\centering Infrastructure network connecting centroids by road and rail.]{
    \includegraphics[height=0.2\paperheight]{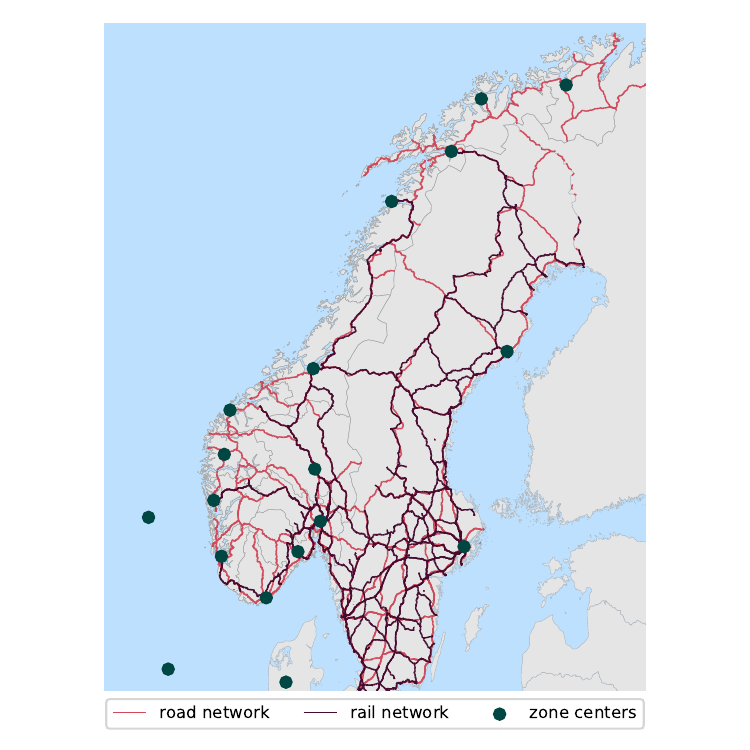}
    \label{fig:zones_infrastructure}
    }%
    \subfloat[\centering Final topology of STraM-Norway. Dashed lines represent investment options.]{
    \includegraphics[height=0.2\paperheight]{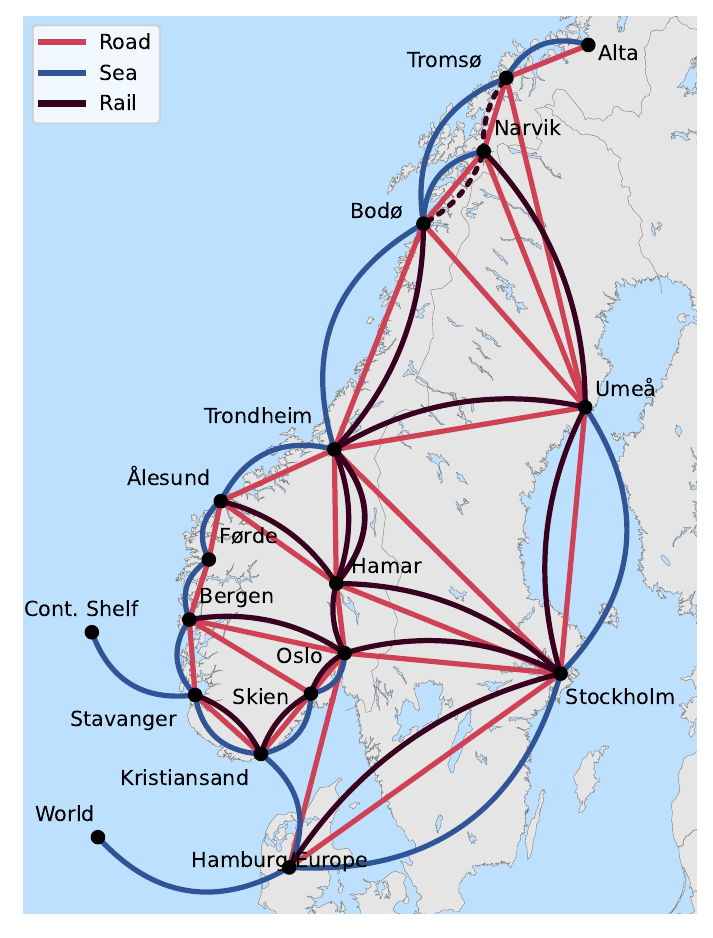}
    \label{fig:topology}
    }%
    \caption{Final configuration of the STraM-Norway topology in 
    \protect\subref{fig:topology} compared with \protect\subref{fig:zones_demand} the demand and \protect\subref{fig:zones_infrastructure} the infrastructure network.}%
    \label{fig:zones}%
\end{figure}

To represent the transportation network in Norway, we divide the country into zones with a centroid (city) and map each centroid to a node in the graph. To get a good graph representation, two criteria should be met: (i) the nodes should accurately map to the origins and destinations of transport demand and (ii) the edges should accurately map to the physical transportation infrastructure. We take the Norwegian counties as a starting point for the zones, with their capital cities as centroids. We also add a zone for the Norwegian continental shelf, northern and southern Sweden, as well as for Europe and the rest of the world, each with a suitable centroid. To achieve criteria (i) and (ii) above, a few adjustments are made based on Figure~\ref{fig:zones_demand} and \ref{fig:zones_infrastructure}, showing the transport demand and physical infrastructure, respectively. The final result is shown in Figure~\ref{fig:topology}.

\subsection{Demand}
\label{Sec:demand}

For the demand of transport, we use detailed data on product movements from \cite{, Madslien.2021}, 
used in the Norwegian National Freight Transport Model (NGM). We aggregate the original 39 product types from NGM into five product groups with similar characteristics: dry bulk, liquid bulk, break bulk, neo bulk, and container. 
A mapping of these original product types into product groups can be found in the online supplementary material. The product groups break bulk and container are further subdivided into two subgroups each: one group of products with a high time urgency (e.g., fresh food) and one with a low time urgency. For the other groups, the time urgency varies little within the groups, so these are not subdivided further. This leaves us with a total of 7 product groups.

Demand is further aggregated by origin and destination zone, using the geographical zones defined in Section~\ref{sec:netwerk_topo}. Demand for transport within zones is neglected, in line with our focus on line-haul transport. The end result is a commodity flow matrix which shows annual demand for transport of products of each group to each pair of origin and destination zone. 

\subsection{Modes, fuels and emissions}
\label{Sec:fuels}

Transport can be performed by three modes: road, rail, and sea. On each mode, a number of fuel technologies are available, some of which are established and mature and some of which are novel and developing. The term ``fuel'' is used in a broad sense and also includes technologies such as batteries (e.g., battery-electric trucks). 

On road, we consider the established fuel diesel and the novel fuels hydrogen and batteries. On rail, we consider the established fuels diesel and electric (catenary trains) and the novel fuels hydrogen and batteries. Finally, on sea we consider the established fuels heavy fuel oil (HFO) and marine gas oil (MGO), and the novel fuels hydrogen, ammonia, and methanol. Current shares of fuels on each mode are based on data from \citet{DNV.2022,DNV.2014,Jernbanedirektoratet.2021}. We estimate $20\%$ of the market share of long-haul road transport to be hard-to-electrify, thus requiring diesel or hydrogen trucks. 

We omit biofuels due to potential limits in production volumes and required land resources \citep{Gray.2021}. Low-carbon technologies like liquid natural gas are excluded since carbon neutrality cannot be reached \citep{Gray.2021}. The availability of renewable fuel technologies increases over time \citep{Gray.2021, Zenith.2019}.

For each fuel, we consider its scope 1 and 2 carbon emissions, considering tailpipe emissions for fossil fuels and emissions during electricity generation for renewable fuels \citep{Martin2023CarbonSectors}. For fossil fuels, emission factors are based on data from \citet{SSB.2021}. For renewable fuels, we assume the fuels are produced from Norwegian grid electricity, and emission factors are based on individual value chain efficiencies from production to consumption, utilizing the model by \citet{Martin2023CarbonSectors}.

\subsection{Transport costs} \label{sec:transp_costs}

Crucial for our model are accurate data regarding transport cost parameters. These costs consist of two parts: transport costs on arcs and transfer costs in nodes.

\begin{figure}[h]
  \centering
  \includegraphics[width=.8\textwidth]{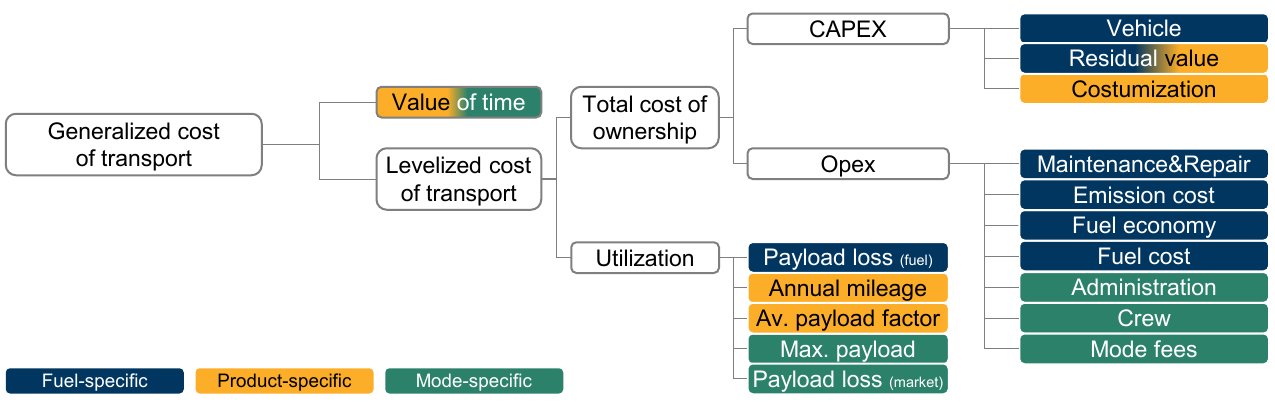}
  \caption{Generalized cost of transport breakdown.
  }
  \label{fig:Cost_structure}
\end{figure}

We compute the \emph{generalized cost of transport} for each mode/fuel/product group combination using an extension of the methodology in \citet{Martin2023CarbonSectors}. The approach decomposes transport cost in its constituent parts (shown in Figure~\ref{fig:Cost_structure}) and estimates all these elements in a harmonized way. This allows for a fair comparison between different modes and fuels, which is crucial in the mode/fuel choice decision in our model. We extend the methodology 
to also include rail freight, distinct product groups, non-monetary costs (the value of time), and improved interpolation between years. 

Besides costs for the physical transportation of the goods, \textit{transfer costs} occur on multimodal paths. Estimated transfer costs at harbors or rail terminals cover non-monetary factors, like the waiting time of cargo in the terminal, as well as the monetary costs of loading and unloading processes and any associated fees, such as port charges.

\subsection{Infrastructure capacities and investments}
\label{Sec:transfer_infrastructure_cost}

Some transport infrastructure in STraM is capacitated. For road freight, we only consider fueling stations as capacity elements, neglecting road capacity limitations. Fueling stations are assumed to be needed every 100 kilometers for battery-electric trucks and every 150 kilometers for hydrogen trucks \citep{Thema.2022}, with associated investment cost based on \citet{TransportandEnvironment.2021}. 

Rail freight faces capacity constraints on railways and in terminals. 
The following investment possibilities are included: single to double track, electrification of non-electrified tracks, and the construction of a new railway along the route Bodø-Narvik-Tromsø (i.e., Nord-Norgebanen), with associated capacities and investment costs based on \citet{Kystverket.2020,jernbaneverket.2015, Aarsland.2021, Jernbanedirektoratet.2021_NordNorgebanen}. 
Capacities of rail terminals and the cost of capacity investments are based on \cite{Kystverket.2020,Grnland.2014,Jernbanedirektoratet.2019_Alnabruterminalen}

Sea freight is only capacitated in terminals (i.e., harbors), given the assumed unlimited traffic volumes at sea. Terminal capacities are based on \cite{SSB.2022_freightvolumeharbour}, while investment costs are extrapolated from Karmsund harbor in Rogaland. 

\subsection{Scenario generation} \label{sec:scen_gen}

The two-stage stochastic programming structure of STraM requires the setup of a scenario tree describing uncertain parameters. Considering too many stochastic parameters can lead to an explosion of the number of scenarios, resulting in computational challenges. Thus, it is crucial to carefully select and define these uncertain parameters. 
We do not expect major changes in economic activities that impact transport demand, nor do we expect average driving speeds across different modes to change substantially. Instead, we account for long-term uncertainty in the cost competitiveness of different fuels. This is the main parameter affecting mode-choice, fuel choice, and investment decisions. 

We construct the scenarios for cost uncertainty as follows. First, we observe that in terms of transport costs, the most significant uncertainty is fuel costs \citep{Martin2023CarbonSectors}. For renewable fuel technologies, we therefore generate scenarios based on the future electricity price as the main cost lever. We generate high/medium/low scenarios for this based on range estimates for Norwegian market zones by \cite{Statnett.2023}. These prices are used as input in the transport cost model from \cite{Martin2023CarbonSectors}, which captures the fact that the cost of different fuels respond differently to electricity price fluctuations. For fossil fuels, the international oil price is the fuels' largest cost lever. Here, we generate high/medium/low scenarios for the oil price based on range estimates by \cite{IEA.2023}, which are used as an index to generate scenarios of all fossil fuel types considered. 
The resulting scenario tree is illustrated in Figure~\ref{fig:scenario_tree}.
\begin{figure}[htbp!]
    \centering
    \scalebox{0.8}{

\def\xd{2}      
\def\yd{0.65}      
\def\xn{-2}      
\def\yn{6}      
\def\vt{.2}      

\begin{tikzpicture}[c/.style 2 args={insert path={node[n={#1}{#2}] (n#1#2){}}},
n/.style 2 args={circle,fill,inner sep=1pt}
]
\draw (\xn,\yn)[c={1}{1}] -- (\xn+1*\xd,\yn)[c={1}{2}] -- (\xn+2*\xd,\yn+3*\yd)[c={1}{3}] -- (\xn+3*\xd,\yn+3*\yd)[c={1}{4}] -- (\xn+4*\xd,\yn+3*\yd) [c={1}{5}] node[right]{HH};
\draw (\xn,\yn)[c={2}{1}] -- (\xn+1*\xd,\yn)[c={2}{2}] -- (\xn+2*\xd,\yn+2*\yd)[c={2}{3}] -- (\xn+3*\xd,\yn+2*\yd)[c={2}{4}] -- (\xn+4*\xd,\yn+2*\yd) [c={2}{5}] node[right]{HM};

\draw[loosely dotted, line width = 2pt] (\xn+2*\xd,\yn+0.5*\yd) -- (\xn+2*\xd,\yn+1.5*\yd);
\draw[loosely dotted, line width = 2pt] (\xn+3*\xd,\yn+0.5*\yd) -- (\xn+3*\xd,\yn+1.5*\yd);
\draw[loosely dotted, line width = 2pt] (\xn+4*\xd,\yn+0.5*\yd) -- (\xn+4*\xd,\yn+1.5*\yd);

\draw (\xn,\yn)[c={3}{1}] -- (\xn+1*\xd,\yn)[c={3}{2}] -- (\xn+2*\xd,\yn)[c={3}{3}] -- (\xn+3*\xd,\yn)[c={3}{4}] -- (\xn+4*\xd,\yn) [c={3}{5}] node[right]{MM};

\draw[loosely dotted, line width = 2pt] (\xn+2*\xd,\yn-.5*\yd) -- (\xn+2*\xd,\yn-1.5*\yd);
\draw[loosely dotted, line width = 2pt] (\xn+3*\xd,\yn-.5*\yd) -- (\xn+3*\xd,\yn-1.5*\yd);
\draw[loosely dotted, line width = 2pt] (\xn+4*\xd,\yn-.5*\yd) -- (\xn+4*\xd,\yn-1.5*\yd);

\draw (\xn,\yn)[c={4}{1}] -- (\xn+1*\xd,\yn)[c={4}{2}] -- (\xn+2*\xd,\yn-2*\yd)[c={4}{3}] -- (\xn+3*\xd,\yn-2*\yd)[c={4}{4}] -- (\xn+4*\xd,\yn-2*\yd) [c={4}{5}] node[right]{LM};
\draw (\xn,\yn)[c={4}{1}] -- (\xn+1*\xd,\yn)[c={4}{2}] -- (\xn+2*\xd,\yn-3*\yd)[c={4}{3}] -- (\xn+3*\xd,\yn-3*\yd)[c={4}{4}] -- (\xn+4*\xd,\yn-3*\yd) [c={4}{5}] node[right]{LL};

\draw (\xn,\yn-4*\yd) -- (\xn+\xd,\yn-4*\yd) -- (\xn+2*\xd,\yn-4*\yd) -- (\xn+3*\xd,\yn-4*\yd) -- (\xn+4*\xd,\yn-4*\yd);
    
\draw (\xn+0*\xd,\yn-4*\yd) -- (\xn+0*\xd,\yn-4*\yd+\vt);
\draw (\xn+1*\xd,\yn-4*\yd) -- (\xn+1*\xd,\yn-4*\yd+\vt);
\draw (\xn+2*\xd,\yn-4*\yd) -- (\xn+2*\xd,\yn-4*\yd+\vt);
\draw (\xn+3*\xd,\yn-4*\yd) -- (\xn+3*\xd,\yn-4*\yd+\vt);
\draw (\xn+4*\xd,\yn-4*\yd) -- (\xn+4*\xd,\yn-4*\yd+\vt);

\node[below] at (\xn+0*\xd,\yn-4*\yd)  {2023};
\node[below] at (\xn+1*\xd,\yn-4*\yd)  {2028};
\node[below] at (\xn+2*\xd,\yn-4*\yd)  {2034};
\node[below] at (\xn+3*\xd,\yn-4*\yd)  {2040};
\node[below] at (\xn+4*\xd,\yn-4*\yd)  {2050};

\end{tikzpicture}
        }
    \caption{Schematic illustration of the scenario tree used in STraM. The scenario names consist of two letters, corresponding to the electricity price and oil price, respectively, indicating whether the price is high (H), medium (M), or low (L) in the scenario.}
    \label{fig:scenario_tree}
\end{figure}
It includes $3^2 = 9$ scenarios, for all combinations of electricity and oil prices. We assume this uncertainty to realize in $2034$, which means that  the periods $2023$ and $2028$ are the first (deterministic) stage, while those from $2034$ to $2050$ are the second (stochastic) stage.

\begin{figure}[htbp!]
    \centering    \includegraphics[width=0.8\linewidth]{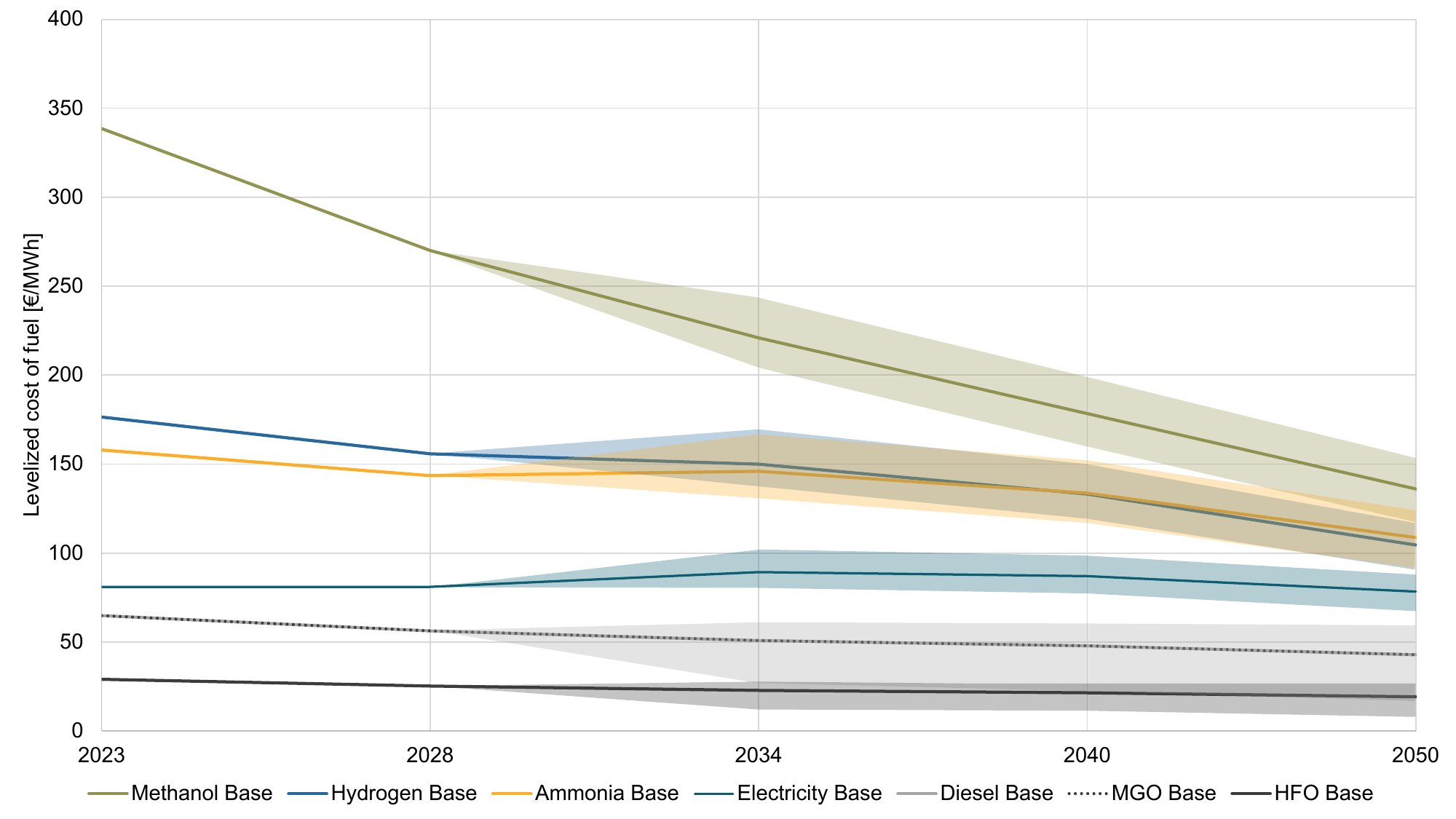}
    \caption{Fuel cost variation due to electricity and oil price uncertainty. The shaded area is generated by using the high and low price scenarios.}
    \label{fig:fuel_scenarios}
\end{figure}

Figure~\ref{fig:fuel_scenarios} shows the impact of these uncertain electricity- and oil prices on fuel costs, which are the driving factor in the calculation of the generalized cost of transport described in Section~\ref{sec:transp_costs}.  
We observe that the impact of electricity price uncertainty correlates with the energy efficiencies of individual fuel value chains. Large fuel cost variations occur in low-efficiency chains, such as those for methanol or diesel. 

\begin{figure}[htbp!]
  \centering
  \includegraphics[width=.8\textwidth]{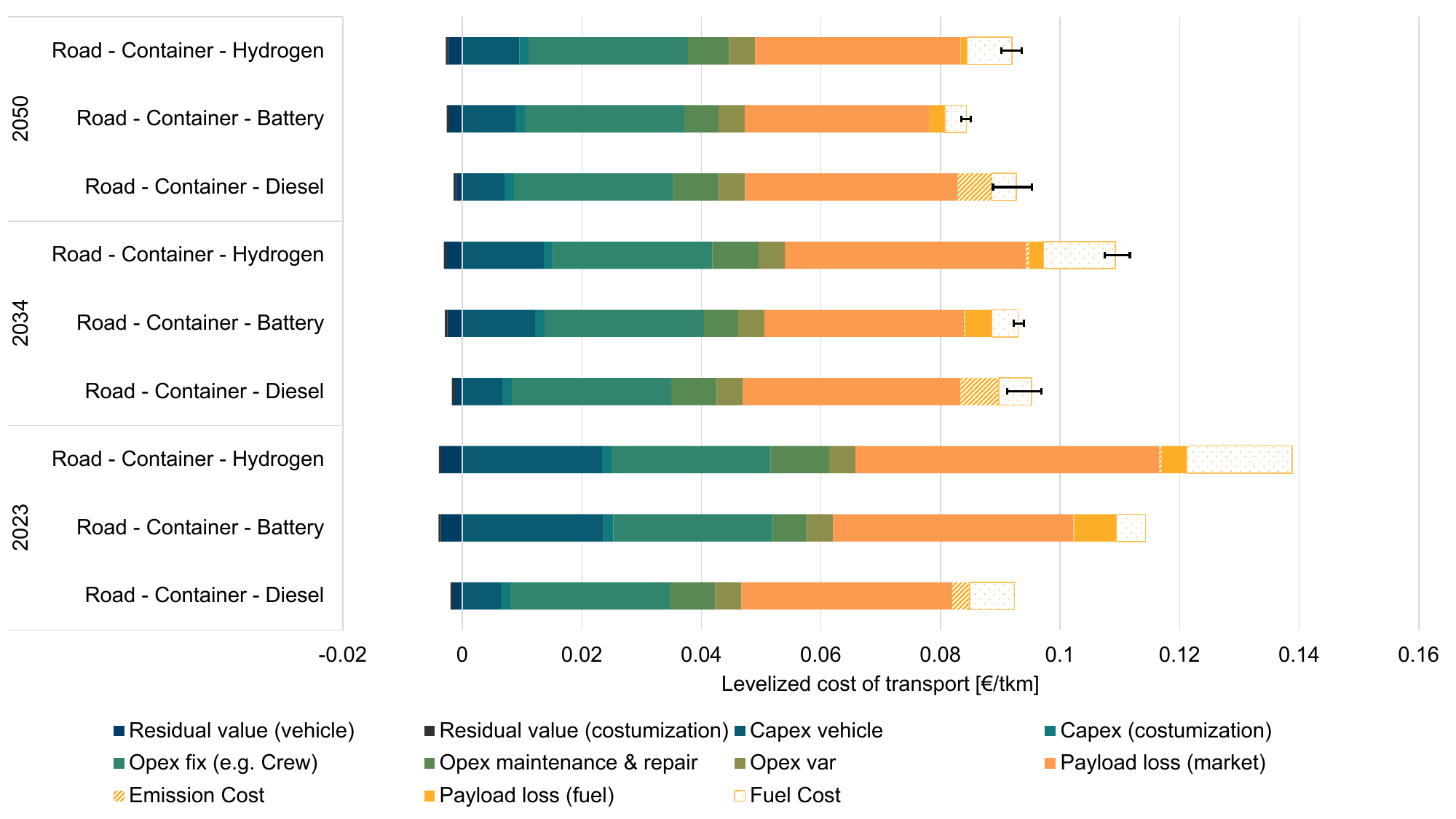}
  \caption{Levelized costs of container transport in line-haul road freight for several fuel technologies and years. Note that negative costs show the vehicle's levelized residual values. Error bars show the electricity and oil price uncertainty for renewable and fossil fuels, respectively.}
  \label{fig:Cost_LCOT_time}
\end{figure}

Finally, Figure~\ref{fig:Cost_LCOT_time} illustrates the impact of fuel uncertainty on the levelized cost of transport, being the main constituent of the generalized cost of transport, for the example of road container transport.
We observe a cost reduction over time for renewable for battery and hydrogen fuels, while diesel costs show a slight increase due to rising carbon prices. Battery-electric trucks become more cost-effective than diesel trucks before 2034, while hydrogen trucks become cost-competitive by 2050. 
\subsection{Path generation}

Our path-flow model formulation requires the a priori generation of a set of admissible paths for each origin-destination pair $(o,d)$. On the one hand, this set should be flexible enough, in the sense that it allows for many different choices of mode. On the other hand, to keep the size manageable, we should only include ``reasonable'' paths (e.g., no unnecessary detours). We strike a balance between these two criteria as follows. We define a set of admissible \textit{sequences of modes} (e.g., road--sea): all sequences of length at most two. As our model does not include first- and last-mile transport, this is not a very restrictive assumption. Then, for each origin-destination pair $(o,d)$ and for each admissible sequence, we include the cheapest path from $o$ to $d$ using that sequence. This can be done using the ``BuildChain'' module in the Norwegian
National Freight Model System \citep{deJong2013Method3}.

One complication is that the question which path of a given mode sequence is cheapest depends on several factors: the product type transported, the fuel used, and the cost scenario. To account for this, we follow the idea in \cite{Zhang2008ApplicationNetwork} and run the procedure above repeatedly for each product type, fuel, and scenario, and merge all results. It turns out that in the overwhelming majority of cases, the same cheapest path is found, so the resulting total number of paths remains limited.

\section{Results} \label{sec:results}

We now present the results of the case study described in the previous section. Section~\ref{sec:mod_fuel_investm} discusses the key outputs, focusing primarily on the resulting investment decisions and modal and fuel mixes. In Section~\ref{sec:carbon_sensitivity} we analyze the implications of various carbon pricing schemes. Finally, Section~\ref{sec:dyn_stat} evaluates the significance of considering the multi-period aspect, while Section~\ref{sec:value_uncertainty} explores the value of modeling long-term uncertainty.

{
STraM-Norway is implemented in Python 3.10.4, formulated using Pyomo and solved using Gurobi 10.0. The model, and corresponding data, is openly available on \url{https://github.com/ntnuiotenergy/STraM}. The analyses are carried out on a Dell Lattitude 7420 computer with an Intel i5-1145G7, 2.60GHz processor, 16GB RAM,  running Windows 10 and using 1 thread. The extensive form of the base model with 9 scenarios consists of 4,554,245  constraints, 1,833,986  continuous variables and 306 binary variables after pre-solve. It solves to optimality (gap of  0.0000\%) in  approximately 4 minutes.}

\subsection{Modal fuel mixes and investment decisions} \label{sec:mod_fuel_investm}

We first discuss results from the \emph{base case} corresponding to the case study described in Section~\ref{sec:case_study} and risk parameters $\lambda=0.3$ and $\gamma = 0.8$. 

\begin{figure}[htbp]
    \centering
    
    \subfloat[Road]{%
        \includegraphics[width=0.32\linewidth]{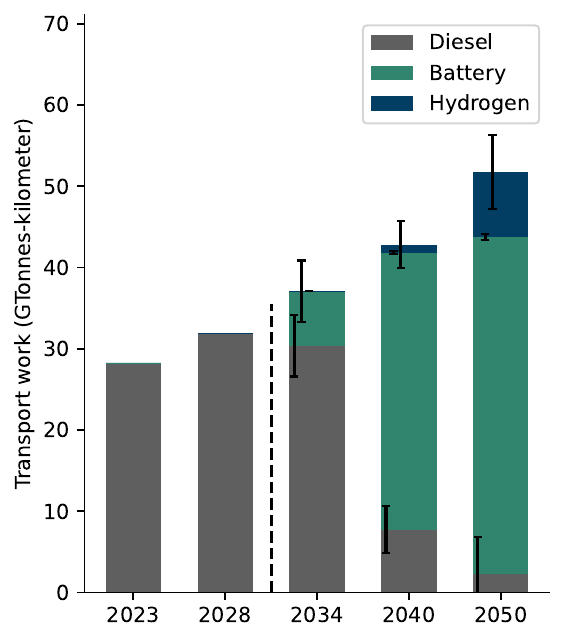} 
        \label{fig:fuel_mix_road}%
    }
    \subfloat[Rail]{%
        \includegraphics[width=0.32\linewidth]{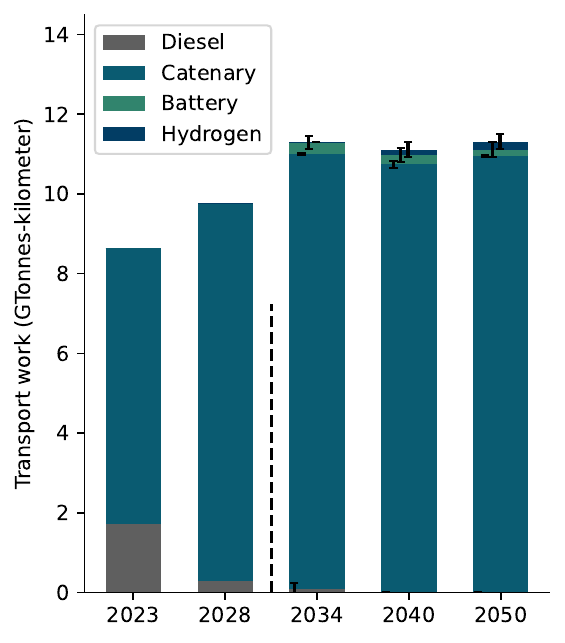} 
        \label{fig:fuel_mix_rail}%
    }
    \subfloat[Sea]{%
        \includegraphics[width=0.32\linewidth]{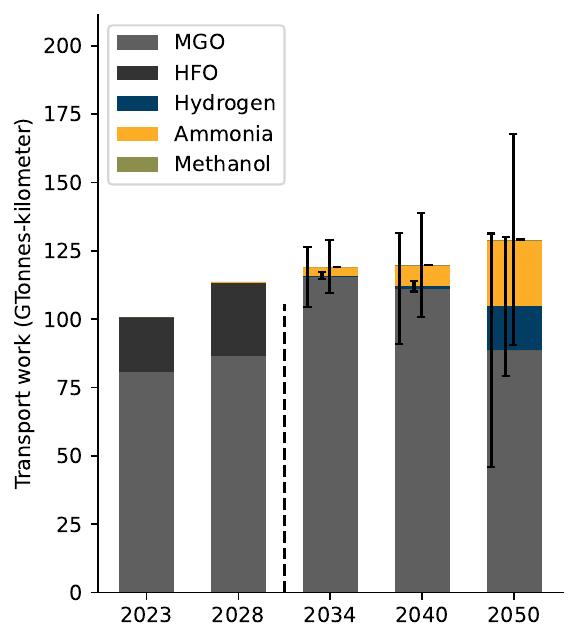} 
        \label{fig:fuel_mix_sea}%
    }
    \caption{Modal fuel mixes for the different transport modes. Standard deviations of second-stage results are visualized using error bars.}
    \label{fig:mode_fuel_mix}
\end{figure}

Figure~\ref{fig:mode_fuel_mix} shows the cost-optimal modal fuel mix. All modes witness an increase in demand with a considerable shift towards road transport, which almost doubles toward 2050. Battery-electric trucks dominate road transport in 2050, with a minor role for hydrogen trucks. On rail, most diesel trains are replaced by catenary trains, but a small role is played by battery and hydrogen trains, which are operated on tracks that are too expensive to electrify. Sea transport shows only a limited shift towards carbon-free fuels, with MGO remaining the dominant fuel in 2050 in most scenarios. Ammonia and Hydrogen take up some market share, depending very much on the fuel cost scenario. 

\begin{figure}[htbp]
    \centering
    \subfloat[Generalized transport costs]{%
        \includegraphics[width=.32\linewidth]{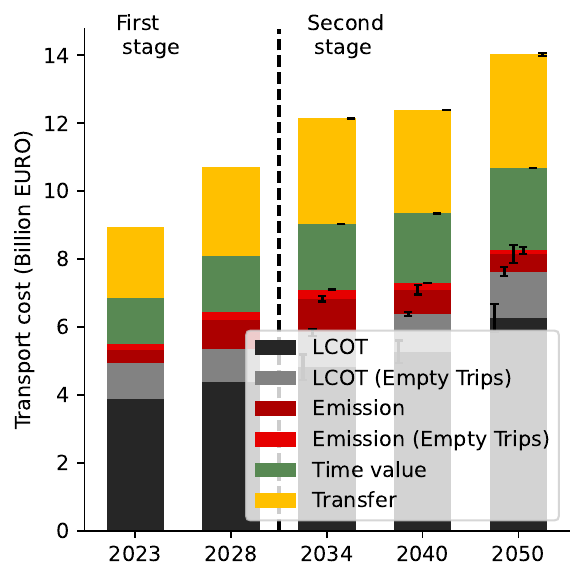} 
        \label{fig:costs_opex}%
    }
    \subfloat[Infrastructure investments]{%
        \includegraphics[width=.32\linewidth]{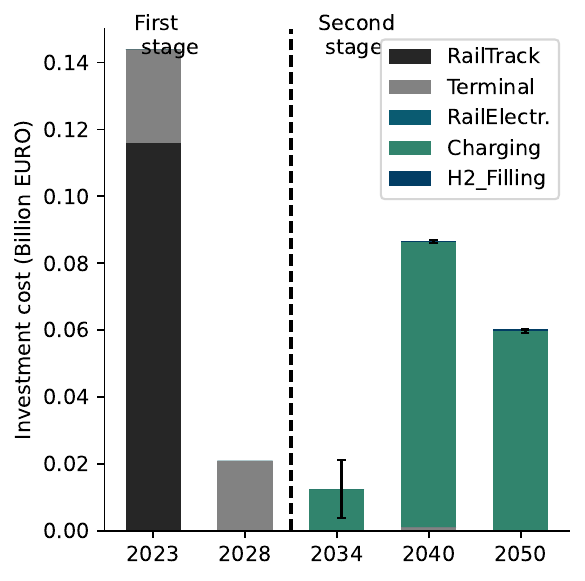} 
        \label{fig:costs_investment}%
    }
    \caption{Cost structures for (\protect\subref{fig:costs_opex}) transport and (\protect\subref{fig:costs_investment}) investments, where the carbon costs are extracted from the generalized transport costs. Implied emissions (relative to the base emissions in 2023) are presented in (\protect\subref{fig:emissions}). The error bars denote standard deviations.}
    \label{fig:results_base}
\end{figure}

\begin{figure}[htbp]
    \centering
    \subfloat[Rail expansion]{%
        \includegraphics[width=.32\linewidth]{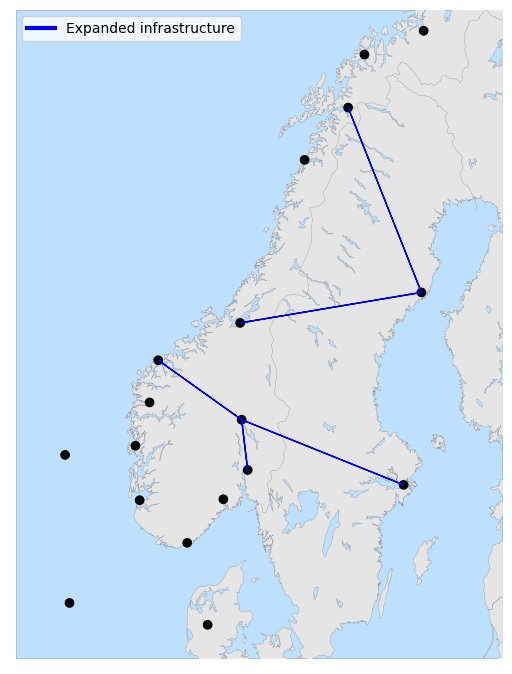} 
        \label{fig:edge_exp}%
    }
    \subfloat[Rail terminal expansion]{%
        \includegraphics[width=.32\linewidth]{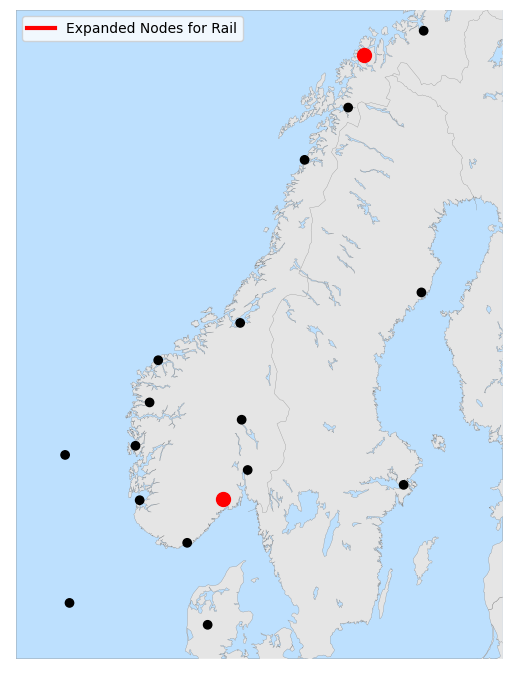} 
        \label{fig:terminal_exp}%
        }\hfill
    \caption{Investments in rail infrastructure in 2023.
    }
    \label{fig:rail_infrastructure}
\end{figure}

\begin{figure}[htbp]
    \centering
    \subfloat[2034]{%
        \includegraphics[width=.32\linewidth]{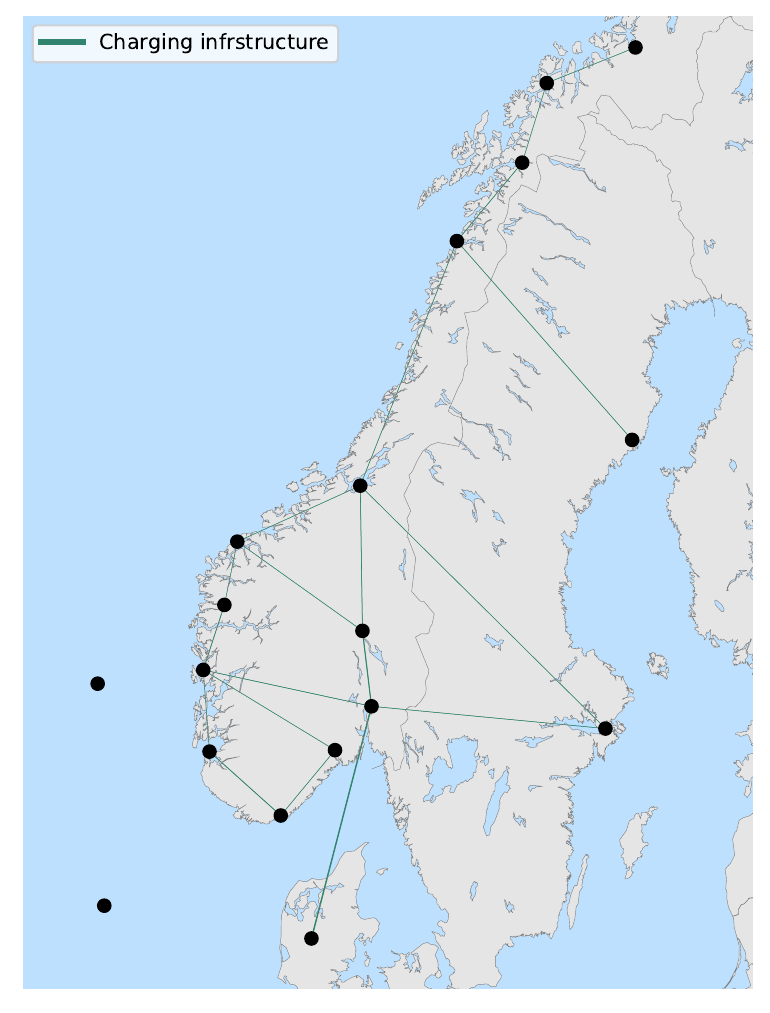} 
        \label{fig:charge2034}%
    }\hfill
    \subfloat[2040]{%
        \includegraphics[width=.32\linewidth]{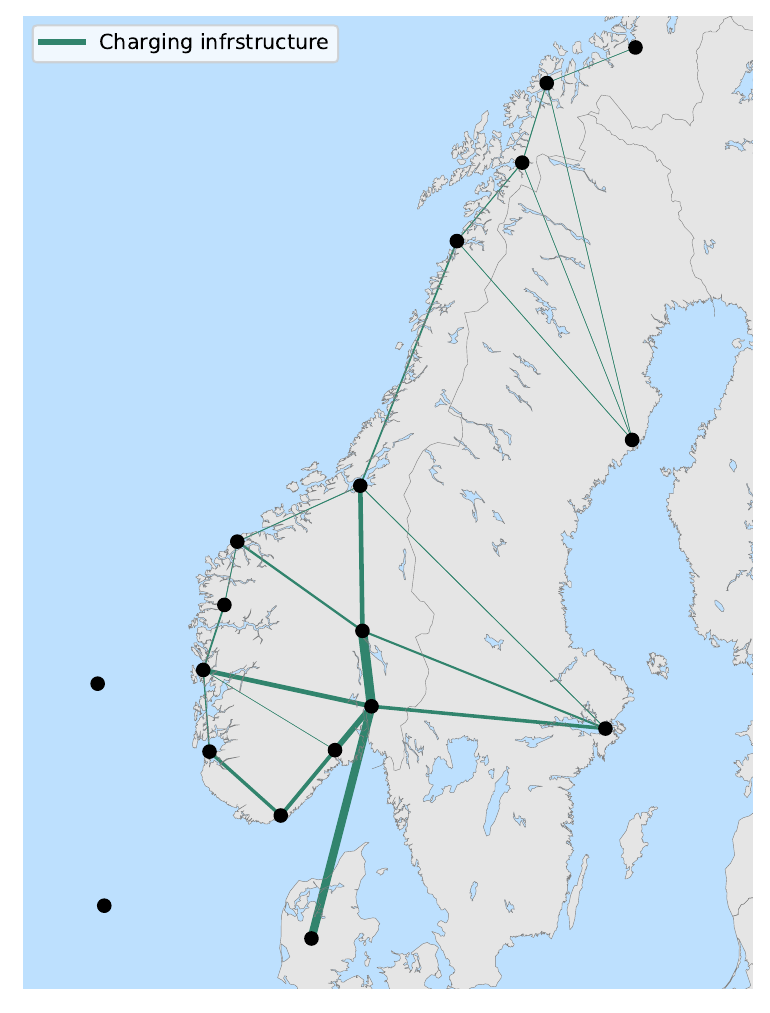} 
        \label{fig:charge2040}%
        }\hfill
    \subfloat[2050]{%
        \includegraphics[width=.32\linewidth]{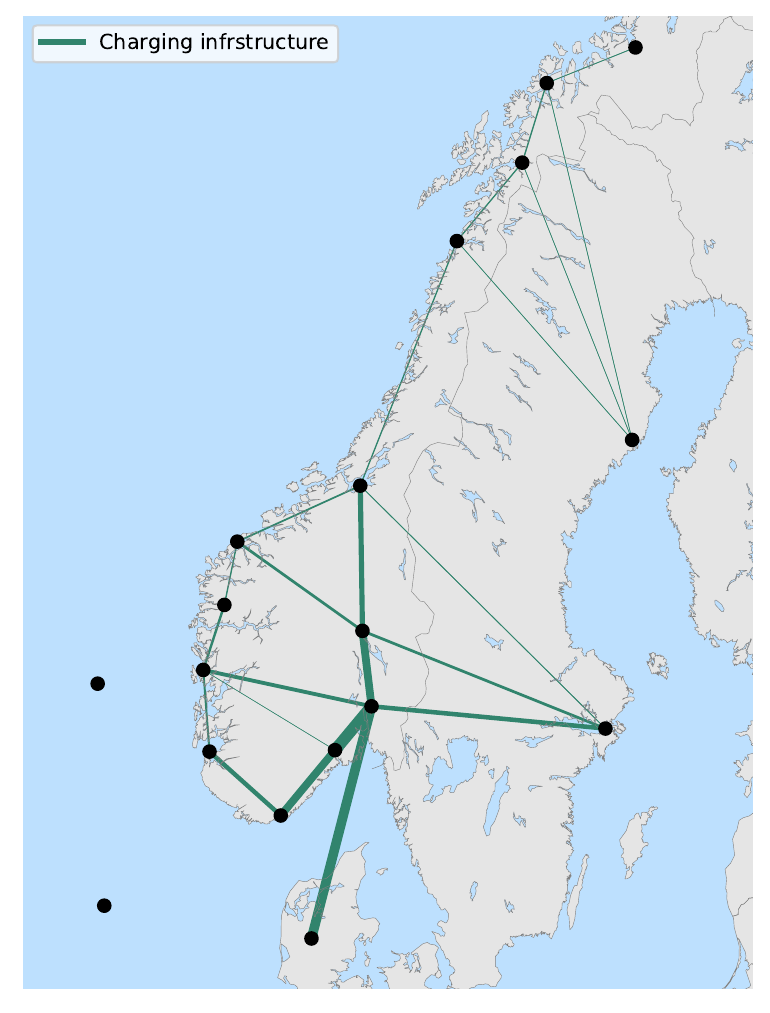} 
        \label{fig:charge2050}%
    }
    \caption{Development of established charging infrastructure over time. The line's thickness illustrates the magnitude of the investment volumes. 
    }
    \label{fig:charging_infrastructure}
\end{figure}

In terms of costs, we can distinguish between transport costs and investment costs, illustrated in Figure~\ref{fig:costs_opex}~and~\ref{fig:costs_investment}, respectively. The total transport costs increase steadily over the years, as a result of increasing demand. The overall cost structure remains stable, with rising emission prices being offset by a decrease in fossil fuel use. Regarding investment costs, we see major investments in the current year. The main driver is investment in rail capacity from Hamar to Oslo, Ålesund and Southern Sweden, as well as from North Sweden to Trondheim and Tromsø, as can be seen from Figure~\ref{fig:edge_exp}. Later years see no additional rail investments, as these projects have long lead times. 
Besides, additional investments in terminals are made, mainly in Oslo, Tromsø and Ålesund, see Figure~\ref{fig:terminal_exp}. In the second-stage, substantial investments are directed towards charging stations for battery-electric trucks and filling stations for hydrogen trucks. As an example, Figure~\ref{fig:charging_infrastructure} illustrates the geographical distribution of the investments in charging infrastructure over time. We observe a gradual increase in charging capacity all over Norway, with higher capacities concentrated along major routes linking the capital Oslo with neighboring areas. 

\subsection{Carbon price analysis} \label{sec:carbon_sensitivity}

\begin{figure}[htbp!]
    \centering
    \includegraphics[width=0.6\linewidth]{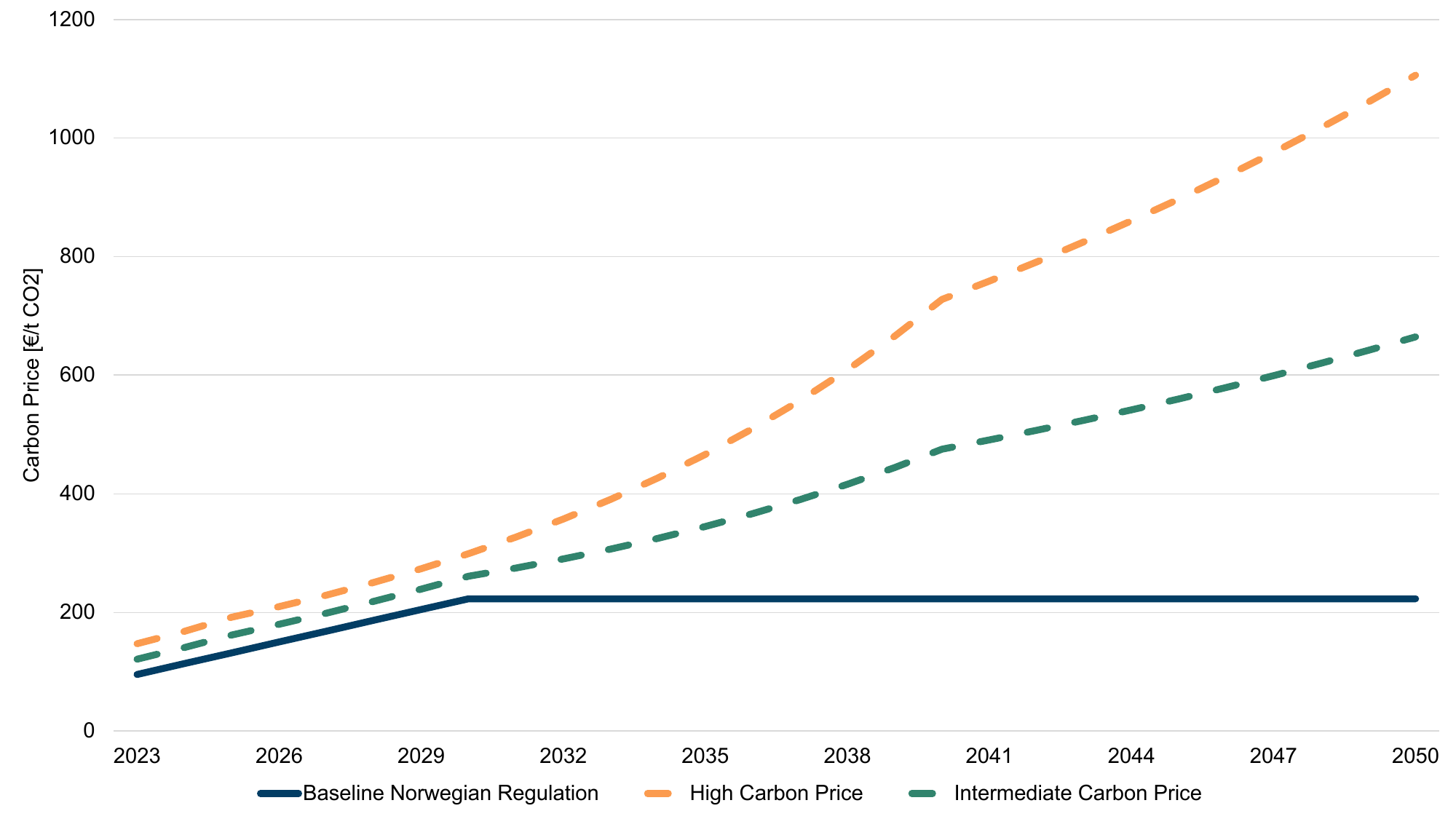}
    \caption{The baseline carbon price scenario, along with an intermediate and high scenario.}
    \label{fig:carbon_price_scenarios}
\end{figure}

To explore the impact of varying carbon policies on emission reduction in the transport system, we conduct a sensitivity analysis.
The \citet{MinistryOfFinance2023} prescribes different carbon price scenarios to be used in techno-economic studies in Norway, and we analyze three of these (base, intermediate, and high), as visualized in Figure~\ref{fig:carbon_price_scenarios}. In addition, we consider a case where we force the system to meet Norwegian emission reductions.

\begin{figure}[htbp]
    \centering
    \subfloat[{\small Baseline}]{%
        \includegraphics[width=.24\linewidth]{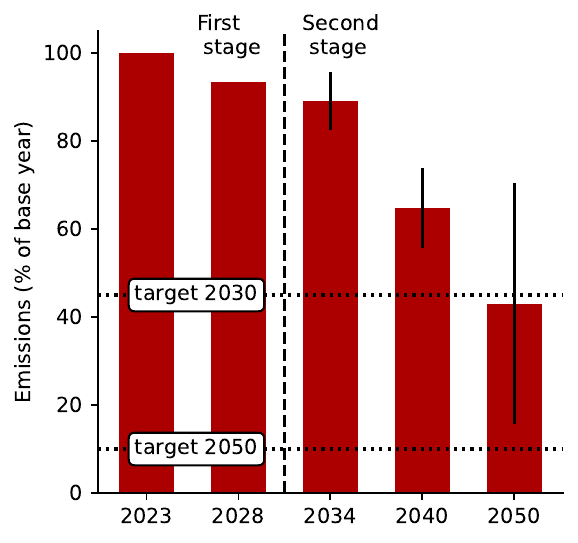} 
        \label{fig:emissions}%
    }\hfill
    \subfloat[{\small Intermediate}]{%
        \includegraphics[width=.24\linewidth]{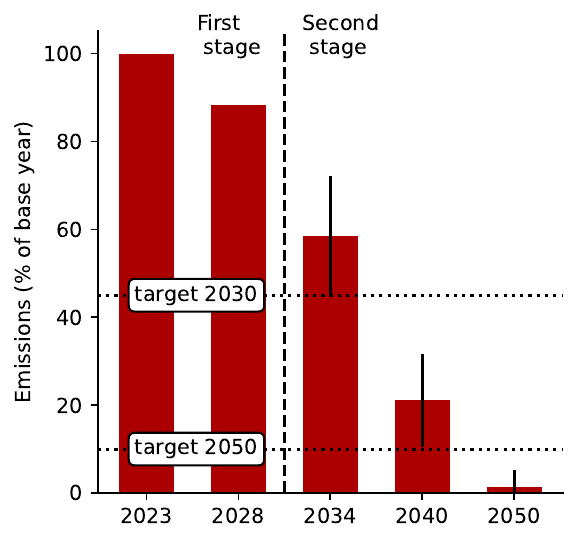} 
        \label{fig:medium_carbon}%
    }\hfill
    \subfloat[{\small High Carbon Price}]{%
        \includegraphics[width=.24\linewidth]{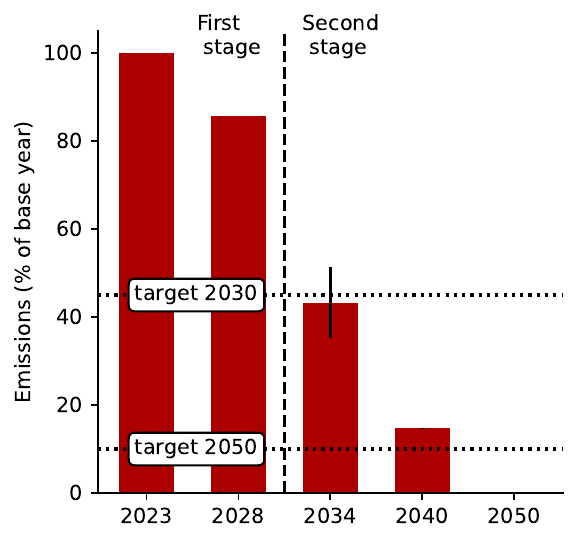} 
        \label{fig:high_carbon}%
        }\hfill
    \subfloat[{\small Forced Emission Red.}]{%
        \includegraphics[width=.24\linewidth]{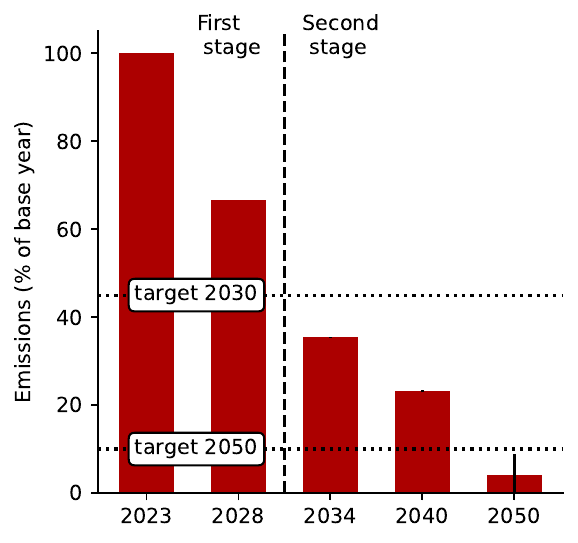} 
        \label{fig:forced_carbon}%
    }
    \caption{Effect of carbon policies on emission reduction: \protect\subref{fig:emissions} baseline carbon price; \protect\subref{fig:medium_carbon} intermediate carbon price  \protect\subref{fig:high_carbon}; high carbon price; and \protect\subref{fig:forced_carbon} system forced to meet Norwegian emission targets. Error bars show the standard deviation caused by second-stage electricity and oil price uncertainty.}
    \label{fig:emission_results}
\end{figure}

Figure~\ref{fig:emissions} plots the total carbon emissions associated with the transport for the base case. By 2050, we see a decrease to approximately 40\% of the emissions in 2023. This significantly falls short of the target of 10\% by 2050. Moreover, the variability in the 2050 emissions is very high, driven by the fuel cost uncertainty. Thus, meeting the 2050 emission target seems both unrealistic and highly dependent on fuel cost developments under the current carbon price system.

Figures~\ref{fig:medium_carbon}-\ref{fig:forced_carbon} show that all intensified policies are effective in reaching the 2050 emission target, but only the last achieves the 2030 target. 
The associated system costs are presented in Table~\ref{tab:investment_sum}. The table shows that for the intermediate carbon pricing, the additional system cost to achieve the 2050 target is relatively mild, but the additional cost for reaching the 2030 target in the forced emission reduction case is significant. 
\begin{table}[htbp]
    \centering
    \begin{tabular}{ccc}
        \toprule
        {Emission scenario} & {Total system costs [B\euro]} & Percentage increase \\ \midrule
        Baseline Norwegian Regulation & 235.4 & -\\ 
        Forced Emission Reduction & 253.0 & 7.5\% \\
        Intermediate Carbon Price & 243.0 & 3.2\%\\
        High Carbon Price & 247.7 & 5.2\% \\\bottomrule
    \end{tabular}
    \caption{Total discounted costs for the system transition under various carbon price scenarios.}
    \label{tab:investment_sum}
\end{table}

\subsection{Dynamic vs. static models} \label{sec:dyn_stat}

\begin{figure}[htbp]
\centering 
    \begin{subfigure}[t]{0.45\textwidth}
        \centering
        \includegraphics[width=0.32\linewidth]{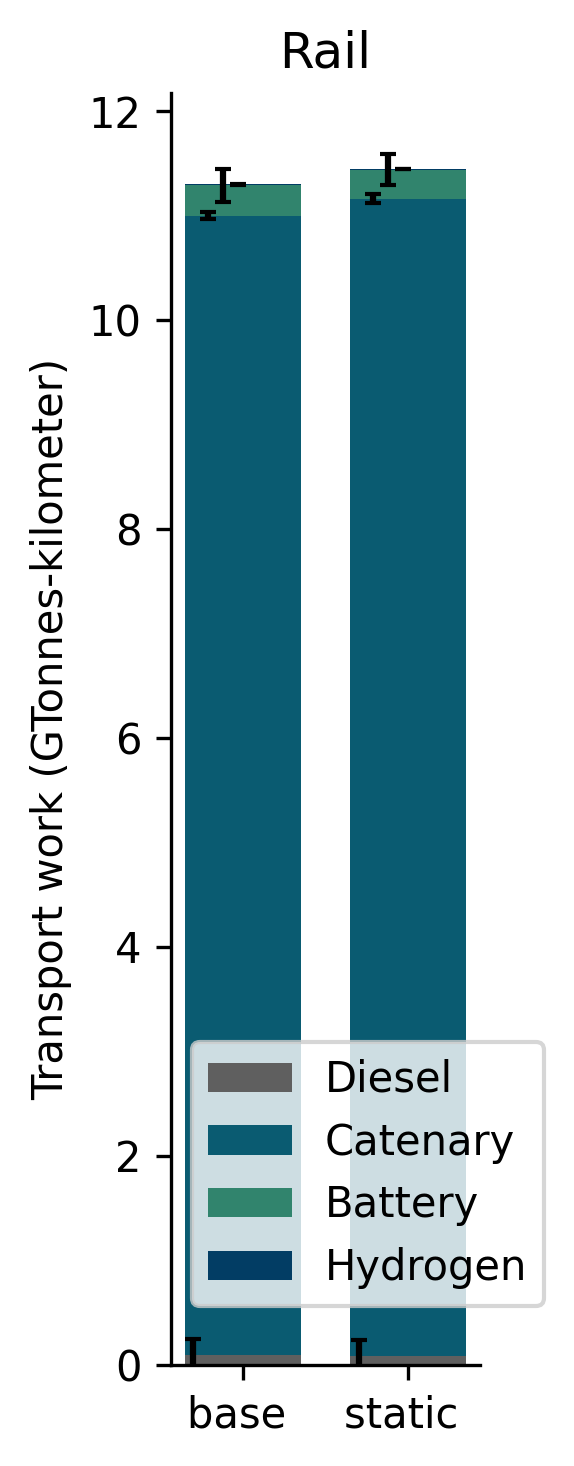} 
        \includegraphics[width=0.32\linewidth]{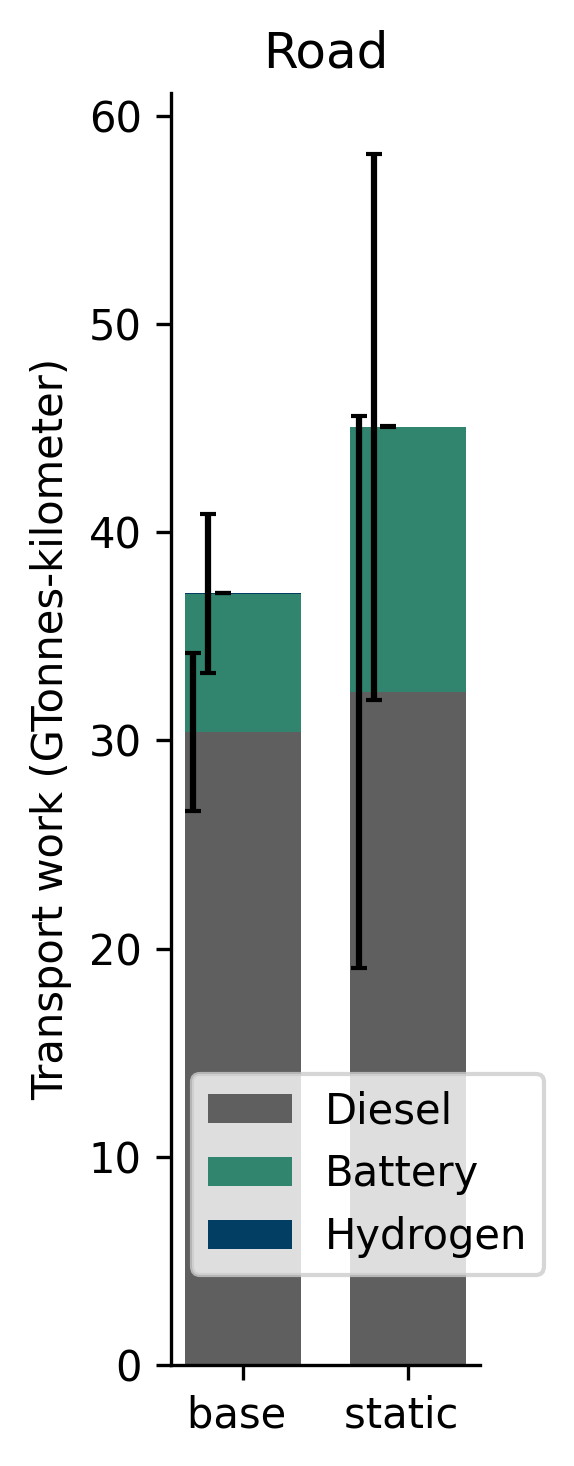} 
        \includegraphics[width=0.32\linewidth]{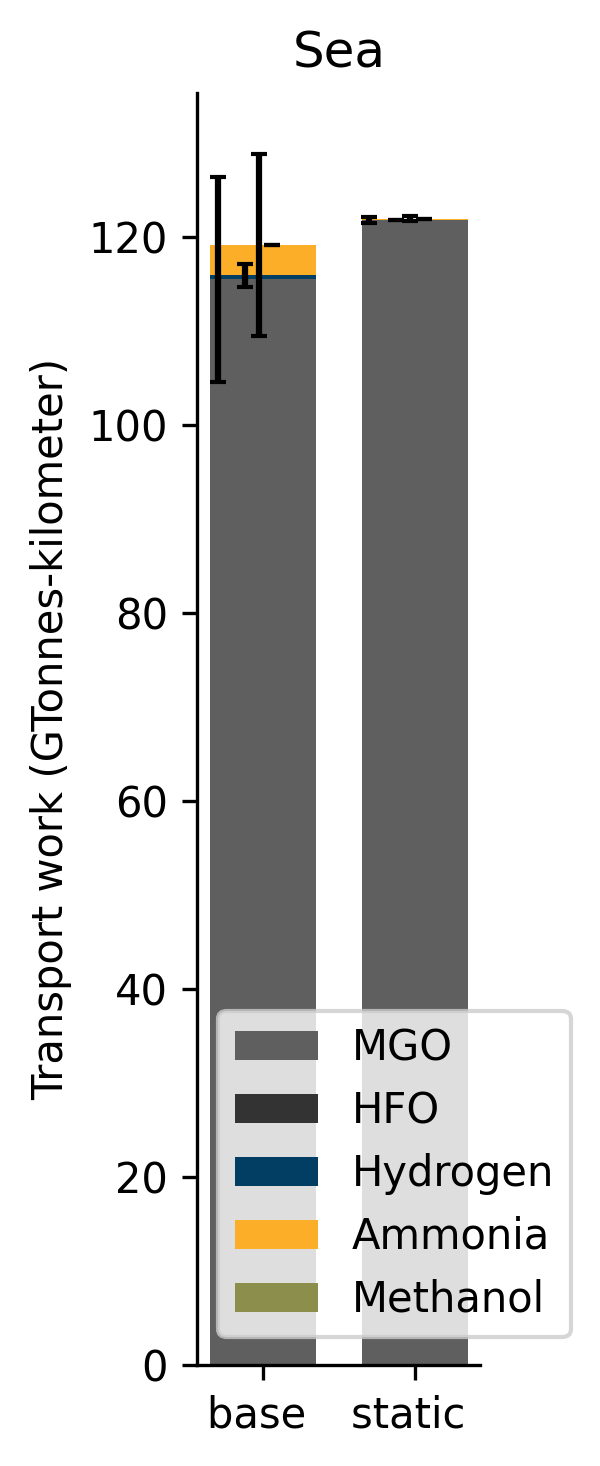} 
        \caption{\textit{Transition} period $2034$} 
        \label{fig:stp_transition}
    \end{subfigure}
    \hspace{0.08\textwidth}
    \begin{subfigure}[t]{0.45\textwidth}
        \centering
        \includegraphics[width=.32\linewidth]{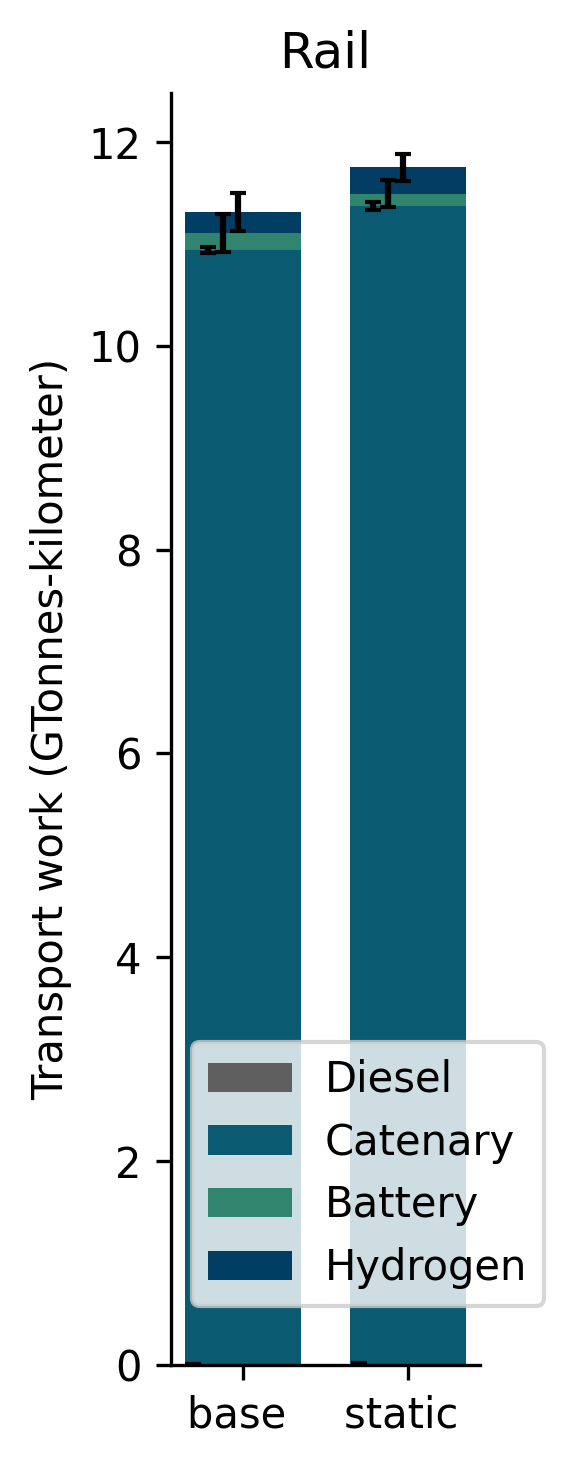} 
        \includegraphics[width=.32\linewidth]{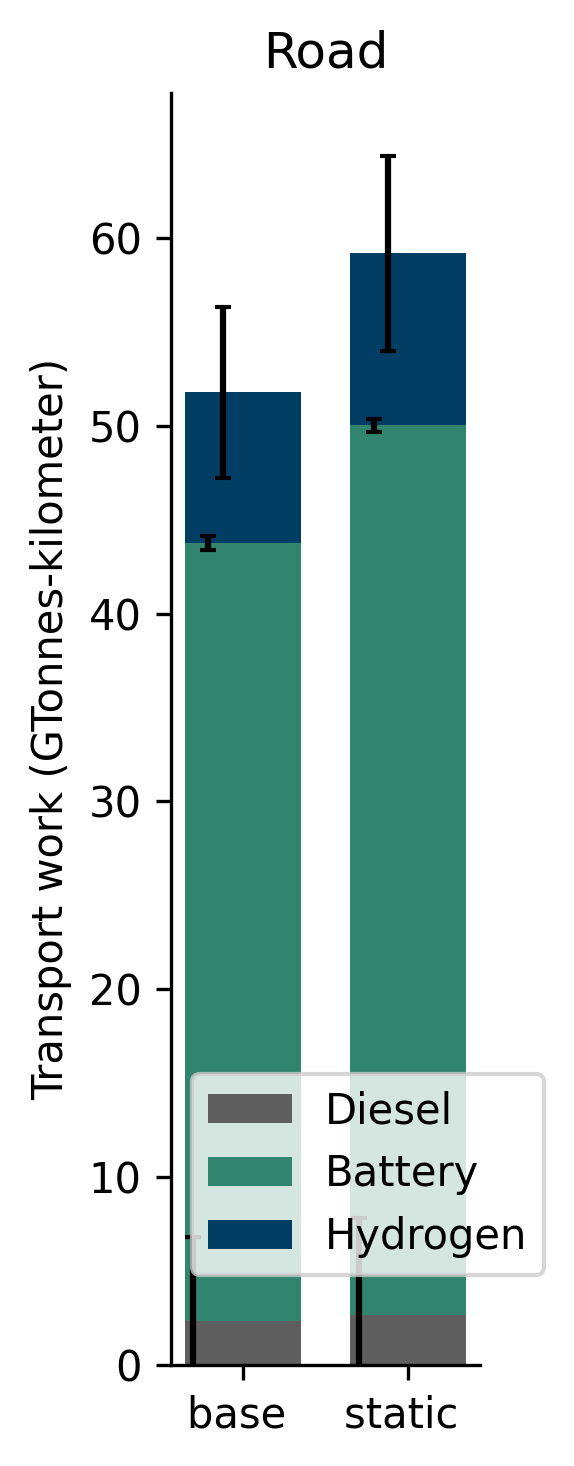} 
        \includegraphics[width=.32\linewidth]{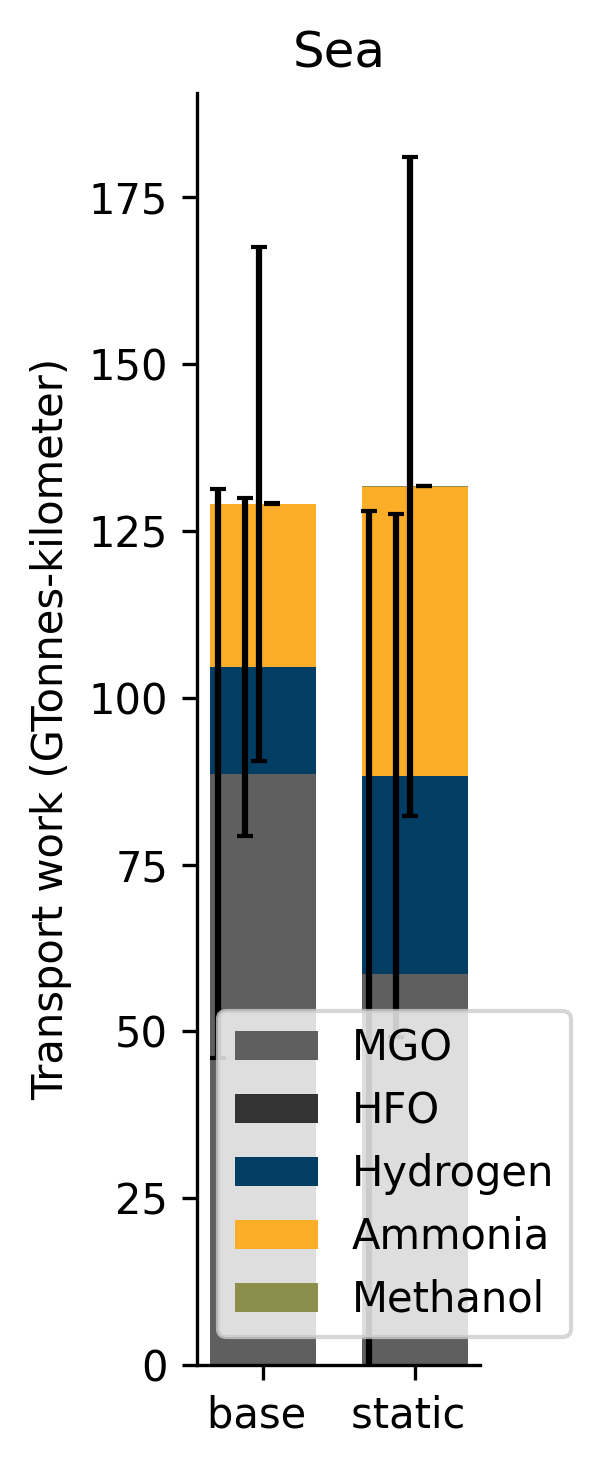} 
        \caption{\textit{Target} period $2050$} \label{fig:stp_final}
    \end{subfigure}
    \caption{Modal fuel mix comparison of the base case with two static cases: a {transition} and {target} period.} 
    \label{fig:single_time_period}
\end{figure}

One of the contributions of STraM is that it explicitly models the development of the transport system over time. To assess the importance of this multi-period aspect, we compare our results with a static version of STraM. 
This static model is constructed by
(i) neglecting the fleet renewal constraints~\eqref{block:fleet_renewal} and technology adaption constraints~\eqref{block:maturity_constraints}, and (ii) only activating the demand, arc-path and fleet balancing constraints~\eqref{block:demand1}-\eqref{block:fleet_balance} in the given \textit{static} time period.
Figure~\ref{fig:single_time_period} compares the modal fuel mix resulting from the full, dynamic version of STraM (indicated as ``base'') with two different static runs. The static run for $2034$ represents a \textit{transition} period, while the static run for $2050$  can be seen as the \textit{target} period.

While the fuel mixes on rail are fairly similar, both in the transition and target period, we observe significant differences in the fuel mixes on road and sea. Considering road transport, we observe that the static case for 2034 allows huge variations in the uptake of battery-electric trucks, freely adjusting to different fuel cost scenarios. The base case from STraM does not show such a huge variation due to the inclusion of, e.g., fleet renewal constraints. For maritime transport, we observe that the base case requires ammonia to be adopted already in the transition period $2034$, to reach the desired market shares in $2050$. The static case does not consider this, and sticks to MGO in $2034$. Thus, STraM is able to give better insights into the dynamic nature of this transition and can sketch a roadmap for how to arrive at a low-carbon transport system in $2050$. Moreover, STraM also provides insight into the timing of the corresponding infrastructure investments. 


\subsection{The value of uncertainty} \label{sec:value_uncertainty}

Another important aspect of STraM is that it explicitly models long-term uncertainty in technology development. We assess the importance of this aspect by comparing the results from our stochastic programming solution (SP) in the base case with the so-called expectation of the expected value solution (EEV).
The EEV is the expected objective value associated with the expected value solution (EV), which is the solution obtained from fixing all uncertain parameters at their expected values. Thus, the EEV represents the expected performance of a solution based on a model that does not explicitly model uncertainty. 

We solved both the SP and EV model and find an objective value of \euro$235.39$ Billion (B) (total system cost) for the SP model, and an EEV of \euro$272.53$B. This means that the value of the stochastic solution (VSS), defined as the difference between the EEV and SP objective is equal to \euro$37.14$B, i.e., the SP solution improves upon the EV solution by $13.6\%$. {In addition, the average total emissions (across all scenarios) are 
$1.9$ Mega Tonnes CO\textsubscript{2} lower for SP than EEV ($6.3\%$). Thus, our stochastic programming approach can help reduce both total costs and emissions, compared to a deterministic approach.}
\section{Discussion} \label{sec:discussion}

In our results, road freight is dominated by battery-electric trucks in 2050 across all scenarios, reaching cost-competitiveness compared to diesel before 2034. The 20\% market share of long-haul road transport that we assume hard-to-electrify is served by either hydrogen or diesel. The choice between the two depends heavily on if and when hydrogen becomes more cost-effective than diesel. Investments in hydrogen filling stations are postponed accordingly. Policymakers might want to hedge the activities of hydrogen truck and infrastructure providers to bridge the gap until full market adoption of hydrogen, potentially supported by synergies with other sectors.

For rail freight, capacity expansion of existing connections with catenary lines is the primary investment strategy. Full electrification of tracks now operated by diesel trains is too expensive; instead they are replaced by battery-electric trains or potentially hydrogen trains, depending on the relative cost development. 

Sea freight appears the most expensive to decarbonize, showing only a partial shift towards ammonia and hydrogen in the base case. The market share of each fuel is highly variable and depends on the realized cost scenario, because fuel costs dominate in maritime transport. Moreover, our model excludes filling infrastructure costs for maritime transport, so the relative attractiveness of both alternatives are potentially somewhat distorted in our results. Thus, we recommend further 
detailed research to find the most attractive renewable fuel system for maritime transport.

Our findings suggest that the minimum carbon prices regulated by the Norwegian government fall short of meeting 2030 and 2050 decarbonization targets. While rail freight achieves carbon neutrality by fully transitioning away from diesel, road and sea freight still rely on fossil fuels to varying degrees, with the maritime sector posing the highest obstacle to Norway's emission goals. Thus, we recommend policymakers to pay close attention to decarbonization in the maritime sector in particular. 

Finally, we discuss some limitations of our model. First, to incorporate the desired strategic elements, like long-term uncertainty and the adoption of new technologies, we sacrificed some of the more detailed logistics elements implemented in the current state-of-the-art national freight transport models. We believe that this is an acceptable limitation as our focus lies on evaluating long-term investments. Moreover, by explicitly modeling the flow of goods in the system in each time period, we still obtain a good approximation of the performance of the operations. Second, we solve a cost-optimization model from a social planning perspective. While the output can yield valuable insights for policymakers into a socially desirable freight transport system, it does not recognize the variety of actors with diverging preferences, resulting in complicated decision dynamics.  Furthermore, we consider a closed, national system, whereas international freight transport relies on global standardization and infrastructure availability, which could significantly impact national decisions. Lastly, our model considers scope 1 and 2 emissions, but neglects life-cycle emissions across fuel technologies.  

\section{Conclusion} \label{sec:conclusion}

In this article, we presented STraM, a strategic network design model for national freight transport decarbonization. By explicitly accounting for the dynamic nature of long-term planning, as well as the development and adoption of new fuel technologies with corresponding uncertainties, STraM is tailored to identify decarbonization pathways and evaluate the impact of government or industry policies. 

Our case study results, targeting the freight transport sector in Norway, show that STraM is able to provide valuable insights into investment strategies and fuel adoption for different transport modes across space and time.  In addition, the explicit consideration of uncertainty in transport costs and technology development leads to a significant improvement in results. In particular, we identify a value of the stochastic solution of approximately $13.6\%$. An interesting result from a policy-maker's perspective is that carbon pricing proves to be an efficient measure to reach emission targets, although relatively high prices are needed.  Moreover, we find that the commonly used \textit{what-if approach} does not manage to consider technology adaption and fleet renewal, leading to a forecasted transport system that is considerably different from the solution that results from STraM. This clearly illustrates the need for a dynamic, multi-period, transport model. 

Overall, we conclude that all the contributions, i.e., the harmonizing of transport cost data; the consideration of multiple time periods with endogenous infrastructure investments; the description of (new) fuel technology adoption; and the inclusion of long-term cost-uncertainty, contribute to better strategic freight transport modeling. 

In future work, we want to connect STraM to a high-resolution national freight transport model (specifically for the ADA framework) to validate its solutions and obtain more detailed insights. 
Similarly, we wish to investigate the relationship with the energy infrastructure that currently falls outside the scope of the model. Other directions for future work are to improve the model accuracy (by, e.g., including new elements such as retrofitting of vessels) or to speed up computations by developing a tailored solution method (heuristic).
\section*{CRediT authorship contribution statement}
	
    \textbf{Steffen J. Bakker, E. Ruben van Beesten:} Conceptualization, Methodology, Software, Formal Analysis, Visualization, Writing - original draft.
    \textbf{Jonas Martin:} Conceptualization, Methodology, Formal Analysis, Visualization, Data curation, Writing - original draft.
    \textbf{Ingvild Synnøve Brynildsen, Anette Sandvig, Marit Siqveland:} Conceptualization, Methodology, Data curation, Software.
    \textbf{Antonia Golab:} Visualization, Writing - Review \& Editing

\section*{Acknowledgements}
This paper was prepared as a part of the Norwegian Research Center on Mobility Zero Emission Energy Systems (FME MoZEES. p-nr: 257653) and the Norwegian Research Centre for Energy Transition Studies (FME NTRANS, p-nr: 296205), funded by the Research Council of Norway. 

We wish to thank Inger Beate Hovi and Anne Madslien (Norwegian Institute of Transport Economics) for providing data and feedback, and Øystein Ullenberg (Institute for Energy Technology) for useful input on some of the model assumptions.  
Moreover, we thank our project partners for valuable feedback and input on requirements for strategic national freight transport modeling during a sequence of workshops. This includes the Norwegian Environment Agency, the Norwegian Public Roads Administration, the Norwegian Coastal Administration, the Norwegian Railway Directorate and the Norwegian Truck Owners Federation. 


\section*{Declaration of Generative AI and AI-assisted technologies in the writing process}

During the preparation of this work the authors used ChatGPT in order to suggest alternative formulations for individual sentences. After using this tool/service, the authors reviewed and edited the content as needed and take full responsibility for the content of the publication.

\bibliographystyle{model5-names-new}   
\bibliography{references} 

\end{document}